\documentclass[11pt]{iopart}
\usepackage{epsfig}

\newcommand{\pprime}{{\prime\prime}}

\newcommand{\bra}{\langle}
\newcommand{\ket}{\rangle}

\newcommand{\order}{{\mathcal O}}

\newcommand{\be}{\begin{equation}}
\newcommand{\ee}{\end{equation}}
\newcommand{\bea}{\begin{eqnarray}}
\newcommand{\eea}{\end{eqnarray}}
\newcommand{\beastar}{\begin{eqnarray*}}
\newcommand{\eeastar}{\end{eqnarray*}}
\newcommand{\bd}{\begin{displaymath}}
\newcommand{\ed}{\end{displaymath}}

\newcommand{\vsp}{\vspace*{3mm}}

\newcommand{\bc}{\ensuremath{\mathbf{c}}}

\newcommand{\bk}{\ensuremath{\mathbf{k}}}

\newcommand{\blambda}{{\mbox{\boldmath $\lambda$}}}

\newcommand{\btau}{{\mbox{\boldmath $\tau$}}}

\newcommand{\bsigma}{{\mbox{\boldmath $\sigma$}}}

\newcommand{\bOmega}{{\mbox{\boldmath $\Omega$}}}

\newcommand{\kav}{{\bar{k}}}
\newcommand{\notdelta}{\overline{\delta}}

\newcommand{\barcij}{{\bar{c}_{ij}}}

\newcommand{\sk}{{S_z(k)}}
\newcommand{\skp}{{S_z(k')}}

\begin{document}

\title[How sampling affects macroscopic features of complex networks]
{What you see is {\em not} what you get: how sampling affects macroscopic features of biological networks}

\author{A Annibale$^\dag$ and ACC Coolen$^{\dag\ddag\S}$}
\address{$\dag$ ~ Department of Mathematics, King's College London, The Strand,
London WC2R 2LS, UK}
\address{$\ddag$ Randall Division of Cell and Molecular Biophysics,
King's College London, New Hunt's House, London SE1 1UL, UK}
\address{$\S$ London Institute for Mathematical Sciences, 22 South Audley St, London W1K 2NY, UK}

\begin{abstract}
We use mathematical methods from the theory of tailored random graphs 
to study systematically the effects of sampling on topological features of large biological signalling networks. 
Our aim in doing so is to increase our quantitative understanding of 
the relation between true biological networks and the imperfect and often biased samples of these networks 
that are reported in public data repositories and used by biomedical scientists. 
We derive exact explicit formulae for degree distributions and degree correlation kernels of sampled networks, in terms 
of the degree distributions and degree correlation kernels of the underlying true network, 
 for a broad family of sampling protocols that include (un-)biased node and/or link  undersampling as well as (un-)biased link oversampling.
Our predictions are in excellent agreement with numerical simulations. 
\end{abstract}

\ead{alessia.annibale@kcl.ac.uk,ton.coolen@kcl.ac.uk}

\section{Introduction}

Networks are popular simplified representations of complex biological many-variable systems. 
The network representation reduces the complexity of the problem by retaining only information on which pairs of dynamical variables in a given system interact, leading to a graph in which the nodes (or vertices) represent the dynamical variables and the links (or edges) represent interacting pairs. If all interactions are symmetric under interchanging the two  variables concerned, the resulting network is nondirected (as e.g. in protein-protein interaction networks). If some or all are nonsymmetric, the network is directed (as e.g. in gene regulation networks). Present-day biological databases contain protein-protein interaction networks and gene regulation networks of many species, with typically in the order of $N\sim 10^3-10^4$ nodes each,  measured and post-processed by various different techniques and protocols. However, in biology the available experimental techniques do not sample the complete system, but only a finite fraction; for the human protein-protein interaction network this fraction is believed to be presently around $0.5$ \cite{Prasad}. Furthermore, the sampling tends to be biased by which experimental method is used \cite{PLOS}. In order to use the available data wisely and reliably it is vital that we understand in quantitative detail how the topological characteristics of a real network relate to those of a finite (biased or unbiased) random sample of this network. If, for instance, we observe that certain modules appear more often (or less often)  than expected in certain cellular signalling networks, we need to be sure that this is not simply a consequence of imperfect sampling. The first studies of the effects of undersampling on network topologies focused on the relation between true and observed degree distributions, either analytically \cite{Stumpf2005a,Stumpf2005b} or via numerical simulation \cite{Han2005}, and found that undersampling changes qualitatively the shape of the degree distribution. Subsequent studies \cite{Lee2006,Silva2006}, based on numerical simulation, revealed the effects of undersampling on topological features other than the degree distribution, such as clustering coefficients, assortativity, and the occurrency frequencies of local motifs. More recent publications were devoted to sampling of non-biological networks, such as the internet \cite{Viger} and bipartite networks \cite{Solokov}. 
So far all published studies on the effects of sampling have either been based on numerical simulations, or been restricted to the effects of sampling on a network's degree distribution. Moreover, there are only very few studies that considered biased sampling (e.g. \cite{Stumpf2005b}), and none that investigate oversampling. In the present paper we use statistical mechanical methods from the theory of tailored random graphs 
to study systematically the effects of sampling on macroscopic topological features of large networks. 
We extend previous work in several ways. Firstly, we investigate the effect of sampling on macroscopic observables beyond the degree distribution, e.g. the joint degree distribution of connected node pairs from which one calculates quantities such as the assortativity. Secondly,  we do this for both unbiased and biased sampling of either nodes, links or both. Thirdly,  we investigate not only network undersampling, i.e. the implications  of false negatives in the detection of links and/or nodes, but also the effects of oversampling, i.e. the implications of false positives. All our results are obtained analytically, and formulated in terms of explicit equations that express degree distributions and degree correlation kernels of observed networks in terms 
of those of the underlying true networks. We test our analytical predictions against numerical simulations and find excellent agreement. 

\section{Definitions}

\subsection{Networks and sampling protocols} 

We consider non-directed networks or graphs.
Each is defined by a
symmetric matrix $\bc=\{c_{ij}\}$, with $i,j=1\ldots N$ and
with $c_{ij}\in\{0,1\}$ for all $(i,j)$. Nodes $i$ and $j$ are connected  if and only if  $c_{ij}=1$. We exclude self-interactions, i.e. $c_{ii}=0$ for all $i$. The degree $k_i(\bc)$  of a node $i$ is $k_i(\bc)\!=\!\sum_j c_{ij}$, the degree distribution of graph $\bc$ is $p(k|\bc)\!=\!N^{-1}\sum_i \delta_{k,k_i(\bc)}$, and we abbreviate its degree sequence  as $\bk(\bc)=(k_1(\bc),\ldots,k_N(\bc))$.  
Sampling stochastically an $N$-node graph $\bc$ will result in observation of an $N^\prime$ node graph $\bc^\prime$.
The relation between  $\bc^\prime$ and  $\bc$ depends on the details of the sampling process. 
We use random variables $\sigma_i\in\{0,1\}$ to denote whether a true node $i$ is observed, and $\tau_{ij}\in\{0,1\}$ whether a link $(i,j)$ is observed (if nodes $i$ and $j$ are). In studying oversampling $\lambda_{ij}\in\{0,1\}$ will indicate whether an absent link is falsely reported as present.
Thus:
\begin{eqnarray}
{\it node~undersampling\!:}&~~& c_{ij}^\prime=\sigma_i\sigma_j c_{ij}  \nonumber \\
{\it bond~undersampling\!:}&~~& c_{ij}^\prime=\tau_{ij}c_{ij} \nonumber  \\
{\it node~and~bond~undersampling\!:}&~~& c_{ij}^\prime=\sigma_i\sigma_j \tau_{ij}c_{ij}  
\label{eq:samplings} \\
{\it bond~oversampling\!:}&~~& c_{ij}^\prime= c_{ij}+(1\!-\!c_{ij})\lambda_{ij} \nonumber 
\end{eqnarray}
In a biological context, node oversampling (e.g. detecting a nonexistent protein) would be unrealistic, so will not be considered in this paper.
Note that $N^\prime=\sum_{i\leq N}\sigma_i$. 
We take all sampling variables $\bsigma=\{\sigma_i\}$, $\btau=\{\tau_{ij}\}$ and $\blambda=\{\lambda_{ij}\}$ to be distributed independently, with the proviso that $\tau_{ij}=\tau_{ji}$, $\lambda_{ij}=\lambda_{ji}$ and $\lambda_{ii}=0$ (so sampled networks remain nondirected and without self-interactions). In unbiased sampling their probabilities are independent of the site indices; in biased sampling the probabilities will depend on the degrees of the nodes involved. We conclude that 
the different types of sampling under (\ref{eq:samplings}) are all special cases of the following unified process:
\begin{eqnarray}
c_{ij}^\prime&=& 
\sigma_i\sigma_j [\tau_{ij}c_{ij}+(1-c_{ij})\lambda_{ij}]~~~~~~\forall (i<j)
\label{eq:cprime}
\end{eqnarray}
with
\begin{eqnarray}
\hspace*{-10mm}
 P(\bsigma,\btau,\blambda|x,y,z)&=& \prod_i\Big[x(k_i)\delta_{\sigma_i,1}\!+\!(1\!-\!x(k_i))\delta_{\sigma_i,0}\Big].
 \prod_{i<j}\Big[y(k_i,k_j)\delta_{\tau_{ij},1}\!+\!(1\!-\!y(k_i,k_j))\delta_{\tau_{ij},0}\Big]
 \nonumber\\
 &&\hspace*{10mm} \times
\prod_{i<j}\Big[\frac{z(k_i,k_j)}{N}\delta_{\lambda_{ij},1}\!+\!(1\!-\!\frac{z(k_i,.k_j)}{N})\delta_{\lambda_{ij},0}\Big]
\label{eq:samplingstats}
\end{eqnarray}
Here $x(k)\in[0,1]$ gives the likelihood that a node with degree $k$ will be detected, $y(k,k^\prime)\in[0,1]$ the likelihood that a link connecting nodes  with degrees $(k,k^\prime)$ will be detected, and $z(k,k^\prime)/N\in[0,1]$ the likelihood that an absent bond will be falsely reported as present (the latter scales as $N^{-1}$ to retain finite connectivity for large $N$). For unbiased sampling the control parameters in (\ref{eq:samplingstats}) would all be degree-independent, i.e.  $x(k)=x$, $y(k,k^\prime)=y$ and $z(k,k^\prime)=z$. We note that, since nonexisting nodes cannot give false negatives, we may always choose $x(0)=y(0,k)=y(k,0)=0$ for all $k$. Typical choices for biased sampling would be $x(k)=k/k_{\rm max}$ and $y(k,k^\prime)=z(k,k^\prime)=kk^\prime/k_{\rm max}^2$, i.e. high-degree nodes and links connecting high-degree nodes are more likely to be reported.

\subsection{Macroscopic characterisation of network structure} 

To control analytically the topological properties of the networks to which our sampling protocols (\ref{eq:samplings}) are applied, we consider the following maximum entropy ensemble of typical graphs with 
prescribed  degrees and prescribed degree correlations:
\begin{eqnarray}
p(\bc)&=& \frac{1}{Z_N}\Big[\prod_i\delta_{k_i,k_i(\bc)}\Big]\prod_{i<j}\Big[\frac{\overline{k}}{N}\frac{W(k_i,k_j)}{p(k_i)p(k_j)}\delta_{c_{ij},1}
+\Big(1-\frac{\overline{k}}{N}\frac{W(k_i,k_j)}{p(k_i)p(k_j)}\Big)\delta_{c_{ij},0}\Big]
\label{eq:ensemble}
\end{eqnarray}
with $p(k)=N^{-1}\!\sum_i\delta_{k,k_i}$ and $\bar{k}=\sum_k p(k)k$, and with $Z_N$ the appropriate normalisation constant. 
Graphs generated according to (\ref{eq:ensemble}) will have $\bk(\bc)=\bk$, $p(k|\bc)=p(k)$, and $\sum_{\bc}p(\bc)W(k,k^\prime|\bc)=W(k,k^\prime)$,  
where $W(k,k^\prime|\bc)=(N\overline{k})^{-1}\sum_{ij}c_{ij}\delta_{k,k_i}\delta_{k^\prime,k_j}$ is the joint degree distribution of connected node pairs.
Apart from the  information in $\bk$ and $W(k,k)$, the ensemble (\ref{eq:ensemble}) is unbiased;
see \cite{Annibaleetal2009} for derivations of its information-theoretic properties, and \cite{CoolenDeMartinoAnnibale2009} 
for MCMC algorithms via which its graphs can be generated numerically. 
The remainder of this paper is devoted to calculating analytically how in typical large networks, with given degree sequences and given degree correlations (i.e. those generated via (\ref{eq:ensemble}), sampling affects the macroscopic topological characteristics $p(k)$ and $W(k,k^\prime)$.  To be specific, we calculate the following quantities in terms of the sampling characteristics 
$\{x(k),y(k,k^\prime),z(k,k^\prime)\}$:
\begin{eqnarray}
\bar{k}(x,y,z)&=& \lim_{N\to\infty}\sum_{\bc}p(\bc)\Big\bra\frac{\sum_{ij}   c^\prime_{ij}}{\sum_i \sigma_i}\Big\ket_{\!\bsigma,\btau,\blambda}
\\
p(k|x,y,z)&=& \lim_{N\to\infty}\sum_{\bc}p(\bc)\Big\bra
\frac{\sum_{i}\sigma_i\delta_{k,\sum_j c_{ij^\prime}}}{\sum_i \sigma_i} \Big\ket_{\!\bsigma,\btau,\blambda}
\\
W(k,k^\prime|x,y,z)&=& \lim_{N\to\infty}\sum_{\bc}p(\bc)\Big\bra
\frac{\sum_{ij} c^\prime_{ij}\delta_{k,\sum_\ell c^\prime_{i\ell}}
\delta_{k^\prime,\sum_\ell c^\prime_{j\ell}}}
{\sum_{ij}c^\prime_{ij}}\Big\ket_{\!\bsigma,\btau,\blambda}
\end{eqnarray}
with $c_{ij}^\prime$ as defined in (\ref{eq:cprime}). The denominators are simplified trivially, using the independence of the sampling variables and the 
definition of $W(k,k^\prime|\bc)$,
since
\begin{eqnarray}
\frac{1}{N}\sum_{i}\sigma_i&=& \frac{1}{N} \sum_i x(k_i)+\order(N^{-1/2})=\sum_k p(k)x(k)+\order(N^{-1/2})
\\
\frac{1}{N}\sum_{ij}c^\prime_{ij}&=&\frac{1}{N}\sum_{ij}x(k_i)x(k_j)\Big[\frac{z(k_i,k_j)}{N}]+c_{ij}[y(k_i,k_j)-\frac{z(k_i,k_j)}{N}]\Big]+\order(N^{-1/2})\nonumber
\\
&=& \sum_{kk^\prime}x(k)x(k^\prime)\Big\{p(k)p(k^\prime)z(k,k^\prime)
+\bar{k} W(k,k^\prime)y(k,k^\prime)\Big\}
 +\order(N^{-1/2})
\end{eqnarray}
We may therefore write
\begin{eqnarray}
\bar{k}(x,y,z)&=&\frac{ \sum_{qq^\prime}x(q)x(q^\prime)\big[p(q)p(q^\prime)z(q,q^\prime)
+\bar{k} W(q,q^\prime)y(q,q^\prime)\big]}{\sum_{q} p(q)x(q)}
\label{eq:average_degree}
\\
p(k|x,y,z)&=& \frac{\lim_{N\to\infty}\sum_{\bc}p(\bc)\Big\bra
N^{-1}\sum_{i}\sigma_i\delta_{k,\sum_j c_{ij^\prime}}\Big\ket_{\!\bsigma,\btau,\blambda}}{\sum_{q} p(q)x(q)}
\label{eq:P_tofind}
\\
W(k,k^\prime|x,y,z)&=& \frac{\lim_{N\to\infty}\sum_{\bc}p(\bc)\Big\bra
N^{-1}\sum_{ij} c^\prime_{ij}\delta_{k,\sum_\ell c^\prime_{i\ell}}
\delta_{k^\prime,\sum_\ell c^\prime_{j\ell}}\Big\ket_{\!\bsigma,\btau,\blambda}
}
{\bar{k}(x,y,z)\sum_{q} p(q)x(q)}
\label{eq:W_tofind}
\end{eqnarray}

\section{Effects of sampling on degree distributions}

\subsection{Connection between observed degree distributions and degree correlations}

We note that  in the case of biased sampling  the average degree (\ref{eq:average_degree})  in the observed graph will
generally depend not only on the degree distribution of the original graph but also on the latter's degree correlations. 
Hence our decision to use the graph ensemble (\ref{eq:ensemble}) for the present study. 
The observed distributions $p(k|x,y,z)$ and $W(k,k^\prime|x,y,z)$ in (\ref{eq:P_tofind},\ref{eq:W_tofind}) are connected via a simple identity, as are $p(k)$ and $W(k,k^\prime)$  in the original graph $\bc$:
\begin{eqnarray}
W(k|x,y,z)&=&\sum_{k^\prime}W(k,k^\prime|x,y,z)\nonumber
\\
&=&\lim_{N\to\infty}
\frac{k}{\bar{k}(x,y,z)}
\sum_{\bc}p(\bc)\Big\bra
\frac{1}{N}\!\sum_{i} \sigma_i\delta_{k,\sum_\ell c^\prime_{i\ell}}
\Big\ket_{\!\bsigma,\btau,\blambda}
\nonumber
\\
&=& 
\frac{k}
{\bar{k}(x,y,z)} ~p(k|x,y,z)
\label{eq:connection}
\end{eqnarray}
So for large $N$ we need to calculate in principle only $W(k,k^\prime|x,y,z)$, 
as $p(k|x,y,z)$ follows via (\ref{eq:connection}). Alternatively, since for unbiased sampling $p(k|x,y,z)$  can  be found analytically with little effort,
the identity (\ref{eq:connection}) can be used for verifying the result of our calculation of   expression (\ref{eq:W_tofind}). 

\subsection{Degree distribution for unbiased sampling}

\begin{figure}[t]
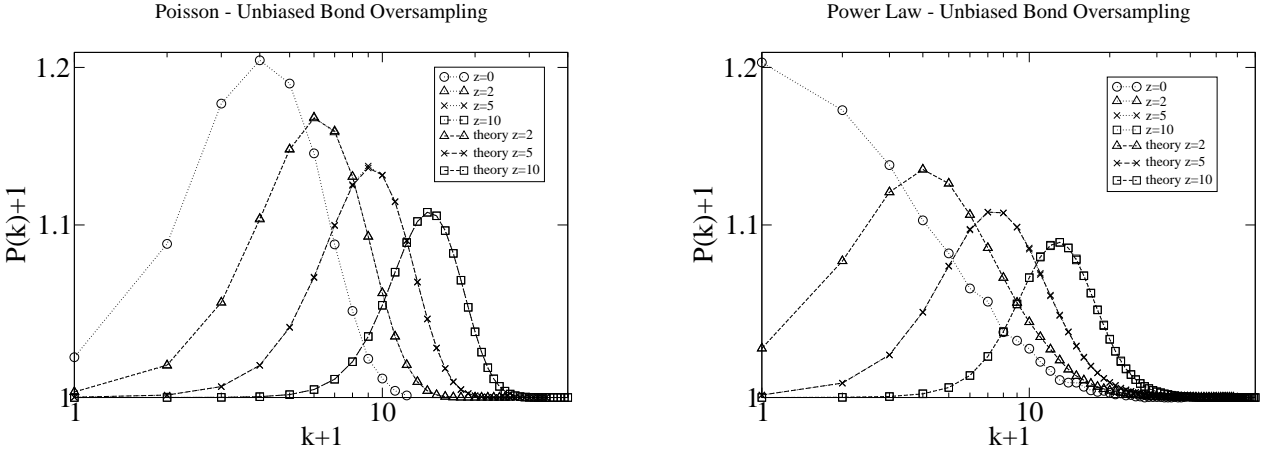

\hspace*{-13mm}
\begin{picture}(200,190)
\put(10,15){\includegraphics[width=215\unitlength,angle=0]{figure/new_poisson_q3.eps}}
\put(270,15){\includegraphics[width=215\unitlength,angle=0]{figure/power_q3.eps}}
\end{picture}
\vspace*{-3mm}
\caption{
Effect of unbiased bond oversampling on the degree distribution of synthetic Poissonian graphs (left panel) and synthetic power law graphs 
(right panel), both with size $N=3512$ and average connectivity $\kav=3.72$.
Different symbols correspond to different fractions $z/N$ of `false positive' bonds, with $z=0,2,5,10$ as shown in the legend. 
The theoretically predicted values (symbols connected by dotted lines) are found to lay perfectly on top of the data points from simulations 
(symbols connected by dashed lines), obtained by averaging over $100$ samples.
}
\label{fig:degree_unbiased_over}
\end{figure}

Calculating $p(k|x,y,z)$ is only straightforward for unbiased sampling, irrespective of whether the source graph is generated according to (\ref{eq:ensemble}), since in that case 
 (\ref{eq:P_tofind}) can be made to factorize over the sampling variables by writing the Kronecker-$\delta$ in integral form. 
In order to appreciate the roles played by the different ingredients of expression (\ref{eq:P_tofind}), we first write it 
 in the form  $p(k|x,y,z)=\lim_{N\to\infty}\sum_{\bc}p(\bc) p_N(k|x,y,z;\bc)$, with 
\begin{eqnarray}
\hspace*{-10mm}&&\hspace*{-15mm}
p_N(k|x,y,z;\bc)=  \frac{1}{\sum_{q} p(q)x(q)\!+\!\order(N^{-1/2})}\frac{1}{N}\sum_{i}\Big\bra
\sigma_i\delta_{k,\sum_j c_{ij^\prime}}\Big\ket_{\!\bsigma,\btau,\blambda}
\nonumber\\
\hspace*{-10mm}
&=&
 \frac{1}{\sum_{q} p(q)x(q)}\frac{1}{N}\sum_{i}
\int_{-\pi}^{\pi}\!\frac{\rmd\omega}{2\pi}\rme^{\rmi k\omega}
\Big\bra
\sigma_i\rme^{-\rmi\omega\sum_j \sigma_i\sigma_j [\tau_{ij}c_{ij}+(1-c_{ij})\lambda_{ij}]}\Big\ket_{\!\bsigma,\btau,\blambda}
+\order(N^{-\frac{1}{2}})
\nonumber
\\
\hspace*{-10mm}
&=&  \frac{1}{\sum_{q} p(q)x(q)}\frac{1}{N}\sum_{i}x(k_i)\!
\int_{-\pi}^{\pi}\!\frac{\rmd\omega}{2\pi}\rme^{\rmi k\omega}\prod_{j\neq i}\Big\{1\!+\!
x(k_j)\Big[
\Big\bra \rme^{-\rmi\omega[\tau_{ij}c_{ij}+(1-c_{ij})\lambda_{ij}]}\Big\ket_{\btau,\blambda}\!\!-\!1\Big]
\Big\}+\order(\frac{1}{\sqrt{N}})
\nonumber
\\
\hspace*{-10mm}
&=& \frac{1}{\sum_{q} p(q)x(q)}\frac{1}{N}\sum_{i}x(k_i)
\!\int_{-\pi}^{\pi}\!\frac{\rmd\omega}{2\pi}\rme^{\rmi k\omega}
\prod_{j\neq i}\Big\{1\!+\!
x(k_j)(\rme^{-\rmi\omega}\!-\!1)\Big[
\frac{z(k_i,k_j)}{N}\!+\!c_{ij}y(k_i,k_j)
\Big]
\Big\}\!+\!\order(\frac{1}{\sqrt{N}})
\nonumber
\\
\hspace*{-10mm}
&=& \frac{1}{\sum_{q} p(q)x(q)}\sum_{q^\prime}x(q^\prime)\!\int_{-\pi}^{\pi}\!\frac{\rmd\omega}{2\pi}\rme^{\rmi k\omega
+(\rme^{-\rmi\omega}-1)\sum_{k^\prime}p(k^\prime)x(k^\prime)
z(q^\prime,k^\prime)}
\frac{1}{N}\sum_{i}\delta_{q^\prime,k_i}
\nonumber
\\
\hspace*{-10mm}&&\hspace*{30mm}\times\rme^{
\sum_{k^\prime}\log\Big\{1+
x(k^\prime)y(q^\prime,k^\prime)(\rme^{-\rmi\omega}\!-1)
\Big\}
\sum_{j}\delta_{k^\prime,k_j}c_{ij}}
+\order(N^{-\frac{1}{2}})
\label{eq:general_p}
\end{eqnarray}
For unbiased sampling protocols, where $x(k)=x$, $y(k,k^\prime)=y$ and $z(k,k^\prime)=z$, this expression immediately simplifies to 
the transparent result
\begin{eqnarray}
\hspace*{-20mm}
p_N(k|x,y,z;\bc)&=& 
\sum_{k^\prime}p(k^\prime|\bc)
\!\int_{-\pi}^{\pi}\!\frac{\rmd\omega}{2\pi}\rme^{\rmi k\omega
+xz(\rme^{-\rmi\omega}-1)}
\Big\{1+
xy(\rme^{-\rmi\omega}\!-\!1)
\Big\}^{k^\prime}
+\order(N^{-\frac{1}{2}})
\nonumber
\\
\hspace*{-20mm}
&=& 
\sum_{k^\prime}p(k^\prime|\bc)\sum_{n\geq 0}\frac{z^n}{n!}
\sum_{m=0}^{k^\prime}\Big(\!\!\begin{array}{c}k^\prime\\ m\end{array}\!\!\Big)x^{n+m}y^m
\!\int_{-\pi}^{\pi}\!\frac{\rmd\omega}{2\pi}\rme^{\rmi k\omega}
(\rme^{-\rmi\omega}\!-\!1)^{n+m}
+\order(N^{-\frac{1}{2}})
\nonumber
\\
\hspace*{-20mm}
&=& 
\sum_{k^\prime}p(k^\prime|\bc)\sum_{n\geq 0}\frac{z^n}{n!}
\sum_{m=0}^{k^\prime}\Big(\!\!\begin{array}{c}k^\prime\\ m\end{array}\!\!\Big)
\Big(\!\!\begin{array}{c}m\!+\!n\\ k\end{array}\!\!\Big)
x^{n+m}y^m
(-1)^{n+m-k} I(k\!\leq\! n\!+\!m)
+\order(N^{-\frac{1}{2}})
\nonumber
\\
\hspace*{-20mm}
&=& x^k
\sum_{k^\prime}p(k^\prime|\bc)\sum_{n\geq 0}\frac{z^n}{n!}
\sum_{m=0}^{k^\prime}\Big(\!\!\begin{array}{c}k^\prime\\ m\end{array}\!\!\Big)
\Big(\!\!\begin{array}{c}m\!+\!n\\ k\end{array}\!\!\Big)
x^{n+m-k}y^m
(-1)^{n+m-k} I(k\!\leq\! n\!+\!m)
\nonumber
\\[-1mm]
\hspace*{-20mm}
&&\hspace*{90mm}+\order(N^{-\frac{1}{2}})
\label{eq:unbiased_p}
\end{eqnarray}
in which $I(S)$ is the indicator function (i.e. $I(S)=1$ if $S$ is true, otherwise $I(S)=0$).
The observed average degree (\ref{eq:average_degree}) for unbiased sampling is, as expected, 
\begin{eqnarray}
\bar{k}(x,y,z)&=& x(z+y\bar{k})
\label{eq:unbiased_kav}
\end{eqnarray}
Formula (\ref{eq:unbiased_p}) simplifies further for various special cases. 
For instance:
\begin{itemize}
\item
{\em Unbiased bond and/or node undersampling}, i.e. $z=0$:
\begin{eqnarray}
p(k|x,y,0)&=& (xy)^k \sum_{k^\prime\geq k}p(k^\prime)\Big(\!\!\begin{array}{c}k^\prime\\ k^\prime\!-\!k\end{array}\!\!\Big)
(1-xy)^{k^\prime-k}
\nonumber
\\
&=& \frac{(xy)^{k}}{k!}\sum_{\ell\geq 0} p(k\!+\!\ell)\frac{(k\!+\!\ell)!}{\ell! }
(1\!-\!xy)^{\ell}
\label{eq:pxy}
\end{eqnarray}
This implies that 
if we sample from a graph with Poissonian degree distribution, i.e.
$p(k)=\kav^k \rme^{-\kav}/k!$, then the degree distribution 
of the sampled graph will be
\bea
p(k|x,y,0)&=&\frac{(xy)^k}{k!}\sum_{k^\prime\geq k}\frac{\kav^{k^\prime} \rme^{-\kav}}{(k^\prime\!-\!k)!}(1\!-\!xy)^{k^\prime-k}
=\frac{(\kav xy)^k \rme^{-\kav xy}}{k!}
\eea
i.e. again a Poissonian distribution, but with a reduced average degree $\bar{k}(x,y,0)= xy\bar{k}$.
This recovers earlier results of \cite{Stumpf2005a,Stumpf2005b}. 
We note also that (\ref{eq:pxy}) is invariant under exchanging $x$ and 
$y$, so  sampling all nodes and a fraction $x=\xi$ 
of the bonds is equivalent to sampling all bonds  and a fraction $y=\xi$ 
of the nodes.
We show in Section \ref{sec:W_unbiased} that this equivalence between bonds and nodes 
under unbiased undersampling also holds for the degree correlations. 
In Figure \ref{fig:degree_unbiased_under} we show the predicted degree distributions 
(\ref{eq:pxy}) together with the corresponding results of numerical simulation of the sampling process, for 
synthetically generated networks with size $N=3512$ and average connectivity $\kav=3.72$ (as in 
the biological protein interaction network of {\em C. Elegans} \cite{Simonis}) and 
Poissonian and power-law degree distributions. The agreement between theory and experiment is perfect. 

 \begin{figure}[t]
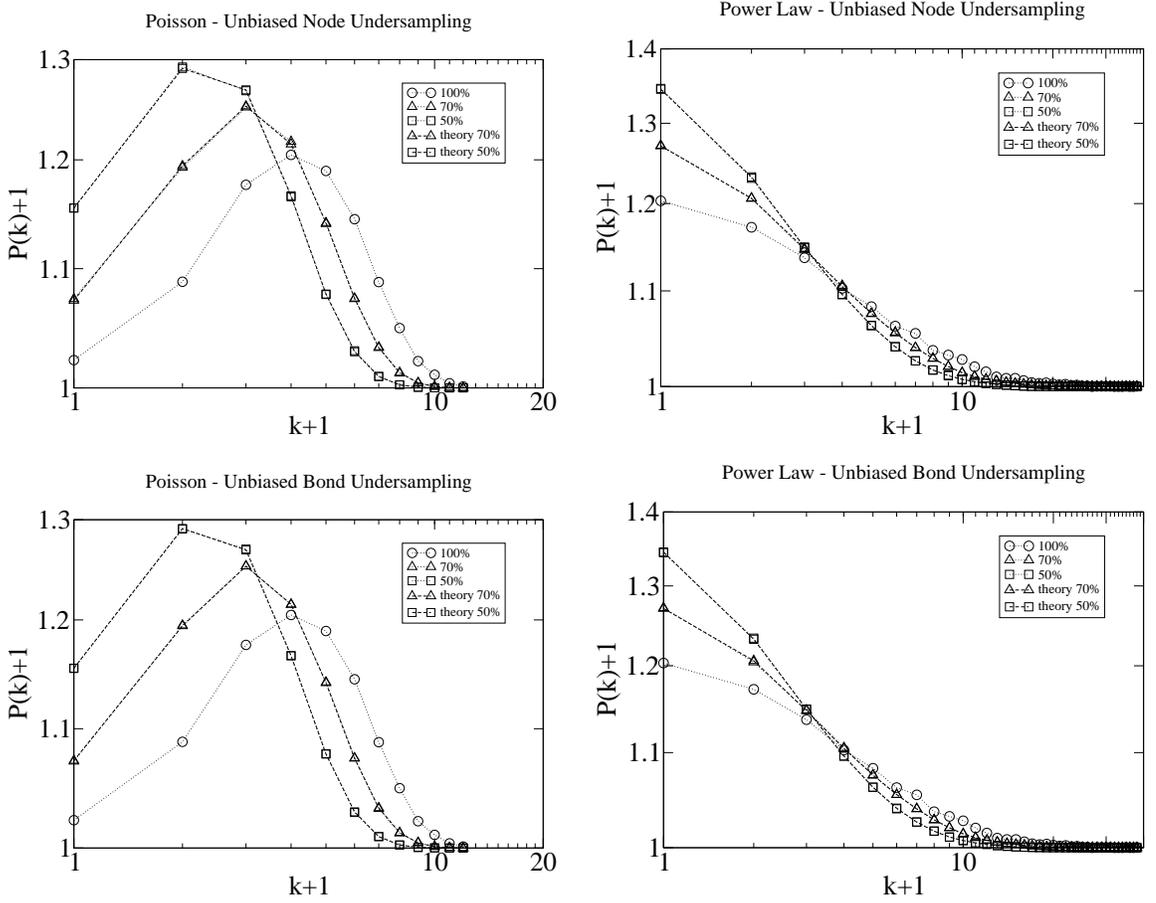

 \unitlength=0.34mm
\hspace*{20mm}
\begin{picture}(200,360)
\put(0,180){\includegraphics[width=215\unitlength,angle=0]{figure/poisson_q1.eps}}
\put(230,180){\includegraphics[width=215\unitlength,angle=0]{figure/power_q1.eps}}
\put(0,0){\includegraphics[width=215\unitlength,angle=0]{figure/poisson_q2.eps}}
\put(230,0){\includegraphics[width=215\unitlength,angle=0]{figure/power_q2.eps}}
\end{picture}
\caption{
Effect of unbiased node undersampling (top panels) and bond undersampling (bottom panels) 
on the degree distribution of synthetically generated networks with size $N=3512$, average connectivity $\kav=3.72$ 
and Poissonian degree distribution (left panels) or power-law distributed degrees (right panels). 
Different symbols correspond to different fractions of sampled nodes ($0.5, 0.7$ and $1$ as shown in the legend) 
and predicted values (symbols connected by dotted lines) lay on top of data points from simulations (symbols connected by dashed lines), 
obtained by averaging over $50$ samples.
}
\label{fig:degree_unbiased_under}
\end{figure}

\item {\em Unbiased bond oversampling}, i.e. $x=y=1$:
\begin{eqnarray}
p(k|1,1,z)&=&
\sum_{k^\prime}p(k^\prime)\sum_{n\geq 0}\frac{z^n}{n!}
\sum_{m=0}^{k^\prime}\Big(\!\!\begin{array}{c}k^\prime\\ m\end{array}\!\!\Big)
\Big(\!\!\begin{array}{c}m\!+\!n\\ k\end{array}\!\!\Big)
(-1)^{n+m-k} I(k\!\leq\! n\!+\!m)
\nonumber
\\
&=& 
\sum_{k^\prime}p(k^\prime)\sum_{n\geq 0}\frac{(-z)^n}{n!}
\sum_{\ell=0}^{n}\Big(\!\!\begin{array}{c}n\\ \ell\end{array}\!\!\Big)
(-1)^\ell\delta_{\ell,k-k^\prime}
\nonumber
\\
&=& \sum_{k^\prime\leq k}p(k^\prime)\sum_{s\geq 0}\frac{z^{k-k^\prime+s}(-1)^s}{s!(k\!-\!k^\prime)!}
=\sum_{\ell=0}^k p(k\!-\!\ell)\rme^{-z}z^\ell/\ell!
\label{eq:p11z}
\end{eqnarray}
As with unbiased undersampling we observe that  sampling from a graph with Poissonian degree distribution, i.e.
$p(k)=\kav^k \rme^{-\kav}/k!$ leads to a sampled graph that is again Poissonian, but 
now with  average degree $\bar{k}(1,1,z)= z+\bar{k}$:
\bea
p(k|1,1,z)&=&
\rme^{-z}\sum_{q\leq k} \frac{\kav^q \rme^{-\kav}}{q!}\frac{z^{k-q}}{(k-q)!}
=\frac{z^k e^{-(\kav+z)}}{k!}\sum_{q\leq k} \frac{k!}{q!(k-q)!}
\left(\frac{\kav}{z}\right)^q
\nonumber\\
&=&\frac{z^k e^{-(\kav+z)}}{k!}\left(1+\frac{\kav}{z}\right)^k
=\frac{e^{-(\kav+z)} (\kav+z)^k}{k!}
\eea
Results from numerical simulations applied to Poissonian and preferential attachment networks are shown in Figure \ref{fig:degree_unbiased_over} 
together with the corresponding theoretical 
predictions. Again 
the agreement between theory and experiment is perfect.
\end{itemize}

\subsection{Degree distribution for biased sampling}

\begin{figure}[t]
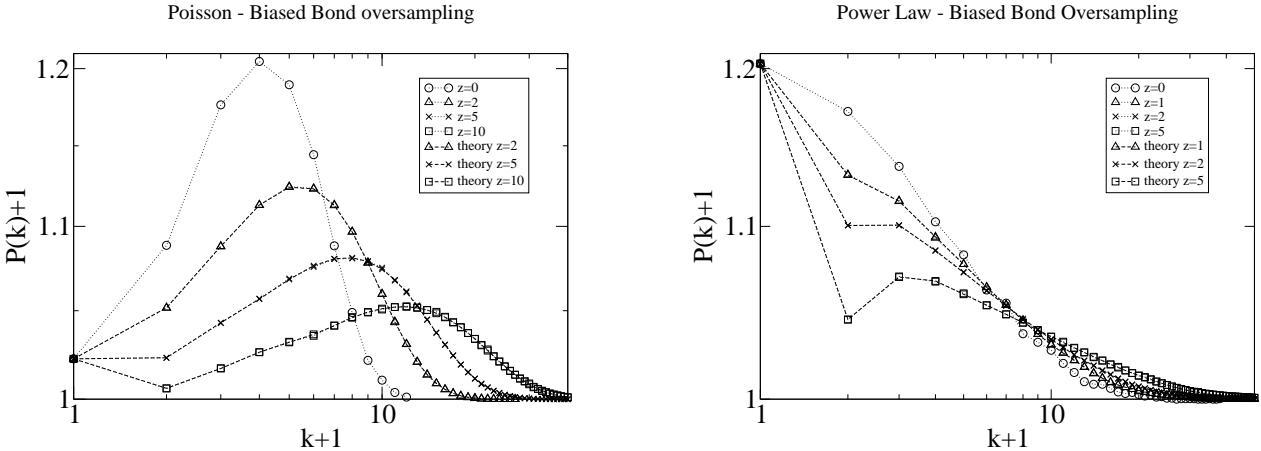

\hspace*{-13mm}
\begin{picture}(200,190)
\put(10,10){\includegraphics[width=215\unitlength,angle=0]{figure/new_poisson_q5.eps}}
\put(270,10){\includegraphics[width=215\unitlength,angle=0]{figure/power_q5.eps}}
\end{picture}\vspace*{-3mm}
\caption{
Effect of biased bond oversampling (i.e. $x(k)=1$, $y(k,k')=1$, $z(k,k')=\alpha k k^\prime/k^2_{\rm max}$) 
on the degree distribution of synthetic Poissonian graphs (left panel) and synthetic power law graphs 
(right panel), both with size $N=3512$ and average connectivity $\kav=3.72$.
Different symbols correspond to different values of $z=\alpha \kav^2/k^2_{\rm max}=0, 2, 5, 10$, as shown in the legend. 
Theoretically predicted values (symbols connected by dotted lines) are found to lay perfectly on top of the data points from simulations 
(symbols connected by dashed lines), obtained by averaging over $100$ samples.
}
\label{fig:degree_biased_over}
\end{figure}

In the case of biased sampling, where $x(k)$, $y(k,k^\prime)$ and $z(k,k^\prime)$ are no longer all degree-independent, 
one can no longer evaluate (\ref{eq:general_p}) without knowledge of the degree-degree correlations in the sources graph $\bc$.
However, the average  (\ref{eq:general_p}) over the graph ensemble  (\ref{eq:general_p}) with controlled degree correlations 
is still feasible. In \ref{app:generalW} we calculate the marginal (\ref{eq:Wmarginals}) of the expected kernel $W(k,k^\prime|x,y,z)$ for the sampled network, from which we obtain 
$p(k|x,y,z)$ via the connection(\ref{eq:connection}). One always has $p(0|x,y,z)=0$, whereas for $k>0$:
\begin{eqnarray}
&&\hspace*{-15mm}
p(k|x,y,z)
= \frac{\sum_qx(q)p(q)\Big\{a(q){\cal J}(k|q)
+qb(q){\cal L}(k|q)\Big\}}
{k\sum_q p(q)x(q)}
\label{eq:general_p}
\end{eqnarray}
with 
\begin{eqnarray}
&&
{\cal J}(k|q)=\rme^{-a(q)}
\sum_{n=0}^{{\rm min}\{k-1,q\}}
\Big(\!\begin{array}{c}q\\ n\end{array}\!\Big)
 \frac{a^{k-1-n}(q)}{(k\!-\!1\!-\!n)!}
b^n(q)(1-b(q))^{q-n}
\label{eq:Jkq}
\\
&&
{\cal L}(k|q)=\rme^{-a(q)}
\sum_{n=0}^{{\rm min}\{k-1,q-1\}}
\Big(\!\begin{array}{c}q-1\\ n\end{array}\!\Big)
 \frac{a^{k-1-n}(q)}{(k\!-\!1\!-\!n)!}
b^n(q)(1-b(q))^{q-1-n}
\label{eq:Lkq}
\\
&& a(q)=
\sum_{q^\prime\geq 0}
p(q^\prime)x(q^\prime)z(q,q^\prime),~~~~~~b(q)=
\frac{\overline{k}}{qp(q)}\sum_{q^\prime\geq 0}
x(q^\prime)y(q,q^\prime)W(q,q^\prime)
\label{eq:aqbq}
\end{eqnarray}
The average connectivity $\bar{k}(x,y,z)$, 
as given in (\ref{eq:average_degree}), is easily obtained from 
(\ref{eq:general_p}) using normalization
of the conditional probabilities ${\cal J}(k|q)$ and ${\cal L}(k|q)$
\begin{eqnarray}
\bar{k}(x,y,z)=\sum_k k\, p(k|x,y,z)=\frac{\sum_q x(q)p(q)[a(q)+q\,b(q)]}{\sum_q p(q)x(q)}
\label{eq:general_kav}
\end{eqnarray}
Let us now work out these results for the `natural' types of sampling bias, where the likelihood of observing nodes or links is proportional to the degrees of the nodes involved, with $\alpha\in[0,1]$:
\begin{itemize}
\item {\em Biased node undersampling}, i.e. $x(k)=\alpha k/k_{\rm max}$, $y(k,k^\prime)=1$, $z(k,k^\prime)=0$:
\\[2mm]
Here we have 
\begin{eqnarray}
&&
a(q)=0, ~~~~~~
q\, b(q){\cal L}(k|q)=k\Big(\!\begin{array}{c}q\\ k\end{array}\!\Big)
b^k(q)(1-b(q))^{q-k}\,I(q\geq k)
\label{eq:no_over1}
\\
&&
\sum_q p(q)x(q)=\frac{\alpha\bar{k}}{k_{\rm max}},~~~~~~ 
b(q)=\frac{\alpha \overline{k}}{qp(q)k_{\rm max}}\sum_{q^\prime>0}
q^\prime W(q,q^\prime)
\label{eq:no_over2}
\end{eqnarray}
This leads to
\begin{eqnarray}
&&\hspace*{-2cm}
p(k|\alpha,1,0)
=\sum_{q\geq k}\frac{qp(q)}{\kav}\Big(\!\begin{array}{c}q\\ k\end{array}\!\Big)
\Big(\frac{\alpha \overline{k}}{qp(q)k_{\rm max}}\sum_{q^\prime>0}
q^\prime W(q,q^\prime)\Big)^k
\Big(\frac{\alpha \overline{k}}{qp(q)k_{\rm max}}\sum_{q^\prime>0}
q^\prime W(q,q^\prime)\Big)^{q-k}
\end{eqnarray}
and
\begin{eqnarray}
\bar{k}(\alpha,1,0)=
\frac{\alpha}{ k_{\rm max}}
 \sum_{qq^\prime>0}qq^\prime W(q,q^\prime)
=\frac{\alpha}{k_{\rm max}}\frac{\overline{k^{(3)}}}{\kav}
\end{eqnarray}
where $\overline{k^{(3)}}=N^{-1}\sum_{ijk\ell}c_{ij}c_{jk}c_{k\ell}$ is the average number of paths of length $3$.
\\[1mm]

\item {\em Biased bond undersampling}, i.e. $x(k)=1$, $y(k,k^\prime)=\alpha kk^\prime/k^2_{\rm max}$, $z(k,k^\prime)=0$:
\\[2mm]
This choice leads again to (\ref{eq:no_over1}), while equations 
(\ref{eq:no_over2}) are now replaced by
\begin{eqnarray}
&&
\sum_q p(q)x(q)=1,~~~~~~ b(q)=
\frac{\alpha\overline{k}}{p(q)k^2_{\rm max}}\sum_{q^\prime> 0}q^\prime 
W(q,q^\prime)
\end{eqnarray}
Hence, one gets
\begin{eqnarray}
&&\hspace*{-2cm}
p(k|1,\alpha,0)
=\sum_{q\geq k}p(q)\Big(\!\begin{array}{c}q\\ k\end{array}\!\Big)
\Big(\frac{\alpha \overline{k}}{p(q)k^2_{\rm max}}\sum_{q^\prime>0}
q^\prime W(q,q^\prime)\Big)^k
\Big(\frac{\alpha \overline{k}}{p(q)k^2_{\rm max}}\sum_{q^\prime>0}
q^\prime W(q,q^\prime)\Big)^{q-k}
\end{eqnarray}
and
\begin{eqnarray}
\bar{k}(1,\alpha,0)=
\frac{\alpha \kav}{ k^2_{\rm max}}
 \sum_{qq^\prime>0}qq^\prime W(q,q^\prime)=\frac{\alpha}{k^2_{\rm max}}\overline{k^{(3)}}
\end{eqnarray}
\\[1mm]

\item {\em Biased bond oversampling}, i.e. $x(k)=y(k,k^\prime)=1$, $z(k,k^\prime)=\alpha kk^\prime/k^2_{\rm max}$:
\\[2mm]
Here we have 
\begin{eqnarray}
&&
a(q)=\sum_{q^\prime \geq 0}
p(q^\prime)z(q,q^\prime)=\frac{\alpha \kav}{k^2_{\rm max}}q,
~~~~~~~~b(q)=1,~~~~~~~\sum_q p(q)x(q)=1\\
&&
{\cal L}(k|q)=e^{-a(q)}\frac{a^{k-q}(q)}{(k-q)!}\,I(q\leq k)\\
&&
{\cal J}(k|q)a(q)=e^{-a(q)}\frac{a^{k-q}(q)}{(k-1-q)!}\,I(q\leq k-1)\equiv (k-q){\cal L}(k|q)
\end{eqnarray}
Substituting into (\ref{eq:general_p}) and (\ref{eq:general_kav}) yields
\be
p(k|1,1,\alpha)=\sum_q p(q){\cal L}(k|q)=\sum_q p(q) e^{-q\alpha \kav/k^2_{\rm max}}\frac{\Big(q\alpha \kav/k^2_{\rm max}\Big)^{k-q}}{(k-q)!}
\label{eq:p_biased_over}
\ee
and
\begin{eqnarray}
\bar{k}(1,1,\alpha)=\bar{k}+\frac{\alpha \kav^2}{k^2_{\rm max}}
\end{eqnarray}
In Figure \ref{fig:degree_biased_over} we show the predicted degree distribution (\ref{eq:p_biased_over}) 
together with the corresponding results from numerical simulations of the biased bond oversampling process.
\end{itemize}

\subsection{Summary}
We have seen that the degree distributions of large sampled networks can be calculated 
and written explicitly in terms of the topological characteristics of the true network, for unbiased and biased under- and oversampling. 
From the resulting equations we can draw the following conclusions:
\begin{itemize}
\item Sampling generally affects the shape of the degree distribution of a network,
with the exception of a Poissonian distribution (as for Erdos-Renyi graphs), 
where the sampled network will only have a rescaled average degree compared to the original.  
\item The degree distribution observed after {\em unbiased} node undersampling of a network is identical to that 
following {\em unbiased} bond undersampling,  for any large 
graph, if the two (node- or bond-) sampling probabilities  are identical. 
\item
In contrast, {\em biased} node undersampling (where the probability of observing a node is proportional to its degree) generally leads to a network with a degree distribution that is
different from the one that would result from {\em biased} bond undersampling (where the probability of observing a bond is proportional to the degrees of the two attached nodes). 
\end{itemize}

\section{Effects of sampling on degree correlation function}

In \ref{app:generalW} we calculate the degree correlation function $W(k,k^\prime|x,y,z)$ 
of large networks 
that are sampled according to the general protocol (\ref{eq:cprime}), 
from graphs generated 
from (\ref{eq:ensemble}). The resulting, expressed in terms of the topological properties $p(k)$ and $W(k,k^\prime)$ of the 
true network, is
\begin{eqnarray}
&&\hspace*{-23mm}
W(k,k^\prime|x,y,z)=\nonumber
\\
&&
\frac{\sum_{q,q^\prime>0}x(q)x(q^\prime)\Big\{p(q)p(q^\prime)z(q,q^\prime){\cal J}(k|q){\cal J}(k^\prime|q^\prime)
+\overline{k}W(q,q^\prime)y(q,q^\prime)
{\cal L}(k|q){\cal L}(k^\prime|q^\prime)\Big\}}
{\bar{k}(x,y,z)\sum_q p(q)x(q)}
\nonumber
\\[-1mm]&&\label{eq:fullW}
\label{eq:general_W}
\end{eqnarray}
with $\overline{k}(x,y,z)$ as given in (\ref{eq:average_degree}), two conditional distributions ${\cal J}(k|q)$ and 
${\cal L}(k|q)$  defined in (\ref{eq:Jkq},\ref{eq:Lkq}),  and with the short-hands $a(q)$ and $  b(q)$ defined in (\ref{eq:aqbq}).
We will now work out this general result for the most common types of sampling, viz. 
node undersampling, bond undersampling, and bond oversampling, including both unbiased and biased protocols.

\subsection{Degree correlations for unbiased sampling}
\label{sec:W_unbiased}
For unbiased sampling protocols where $x(q)=x$, $y(q,q')=y$
and $z(q,q')=z$, one has $a(q)=xz$, $b(q)=xy$ and 
${\cal L}(k|q)={\cal J}(k|q-1)$, so (\ref{eq:general_W})
simplifies immediately to 
\bea
&&
\hspace*{-2cm}
W(k,k^\prime|x,y,z)=
\frac{\sum_{q,q^\prime>0}\Big\{zp(q)p(q^\prime){\cal J}(k|q){\cal J}(k^\prime|q^\prime)
+y\overline{k}W(q,q^\prime)
{\cal J}(k|q\!-\!1){\cal J}(k^\prime|q^\prime\!-\!1)\Big\}}
{z+y\bar{k}}
\label{eq:W_unbiased}
\eea
with
\begin{eqnarray}
&&
{\cal J}(k|q)= 
\rme^{-xz}x^{k-1}
\sum_{n=0}^{{\rm min}\{k-1,q\}}
\Big(\!\begin{array}{c}q\\ n\end{array}\!\Big)
 \frac{z^{k-1-n}(q)}{(k\!-\!1\!-\!n)!}
y^n (1-xy)^{q-n}
\label{eq:J_unbiased}
\end{eqnarray}
Formula (\ref{eq:W_unbiased}) simplifies further for various special cases:

\begin{itemize}
\item {\it Unbiased node and/or bond undersampling}, i.e. $z=0$.\\[2mm]
Here we obtain
\be
J(k|q)=
\left(
\begin{array}{l}
~~q\\
k-1
\end{array}
\right)
(xy)^{k-1}(1-xy)^{q-k+1}\,I(q \geq k-1)
\ee
so equation (\ref{eq:W_unbiased}) reduces to
\begin{eqnarray}
&&\hspace*{-18mm}W(k,k^\prime|x,y,0)=\sum_{q\geq k}\sum_{q'\geq k'}W(q,q')
\left(
\begin{array}{l}
q-1\\
k-1
\end{array}
\right)
\left(
\begin{array}{l}
q'-1\\
k'-1
\end{array}
\right)
(xy)^{k+k'-2}(1-xy)^{q+q'-k-k'}
\label{eq:W_unbiased_under}
\end{eqnarray}
We note that $W(x,y,0)$, like (\ref{eq:pxy}) previously, is symmetric under exchanging $x$ and $y$, 
i.e. node and bond unbiased undersampling lead to the same degree correlations. Therefore the equivalence between the two samplings   
is now fully 
established for large graphs drawn from ensemble (\ref{eq:ensemble}).
\vsp

Equation (\ref{eq:W_unbiased_under}) clearly shows that sampling from graphs in which degree correlations are present 
will generally affect those correlations, even in Poissonian networks, in spite of the fact that there  the degree distribution is only changed via a reduction of the average degree.
Conversely, if we sample from graphs without degree correlations, i.e. for which $W(k,k^\prime)=W(k)W(k^\prime)=
p(k)p(k^\prime)kk^\prime/\overline{k}^2$, equation (\ref{eq:W_unbiased_under}) reveals that the degree correlation function in the sampled graph
factorizes in the product of its marginals as well, i.e. $W(k,k^\prime|x,y,0)=W(k|x,y,0)W(k^\prime|x,y,0)$.
This means that unbiased bond and/or node undersampling from graphs without degree correlations 
does not generate any degree correlations.

\begin{figure}[t]
 \unitlength=0.31mm
\hspace*{-13mm}
\begin{picture}(200,590)
\put(50,400){\includegraphics[width=200\unitlength,angle=0]{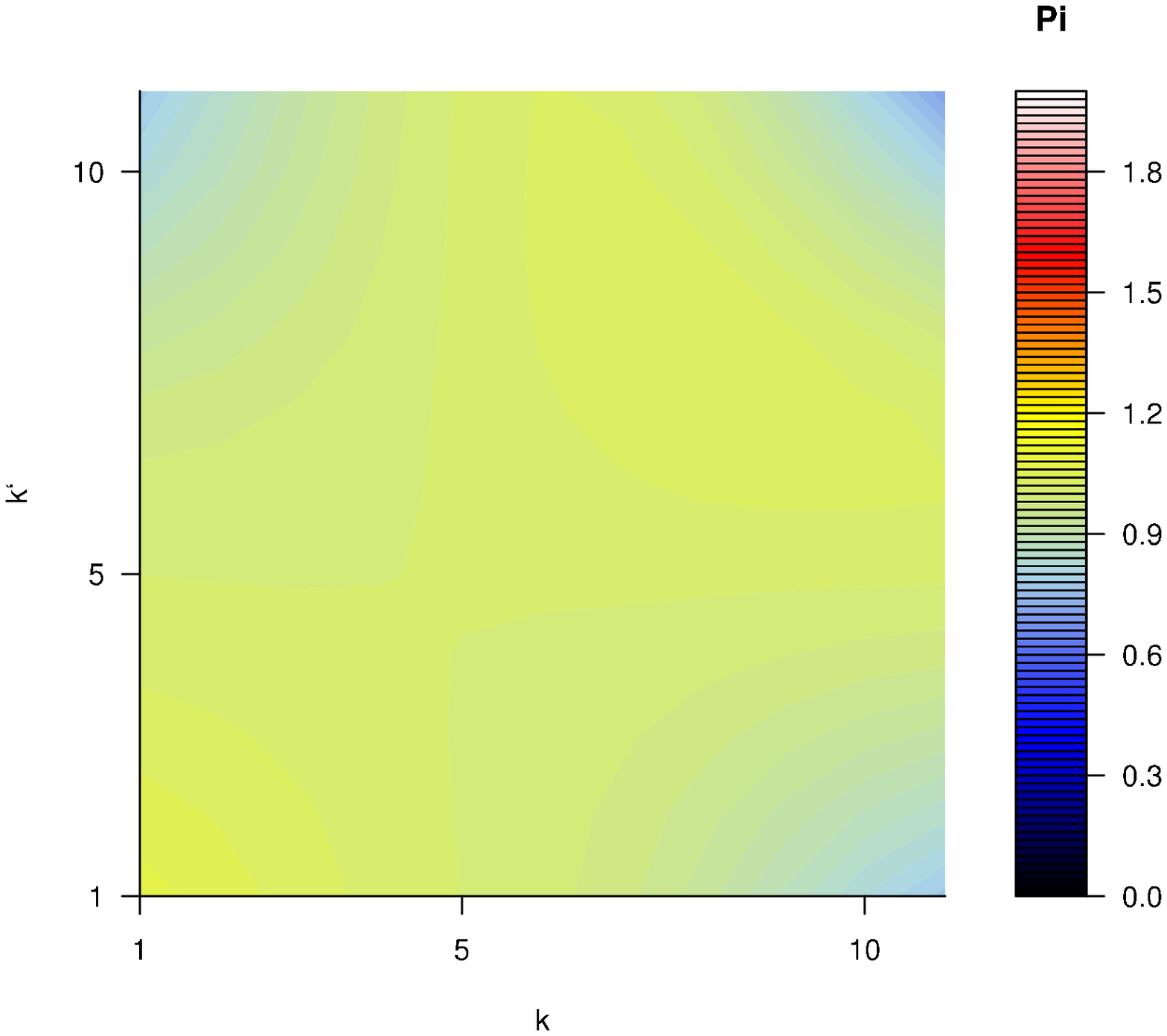}}
\put(270,400){\includegraphics[width=200\unitlength,angle=0]{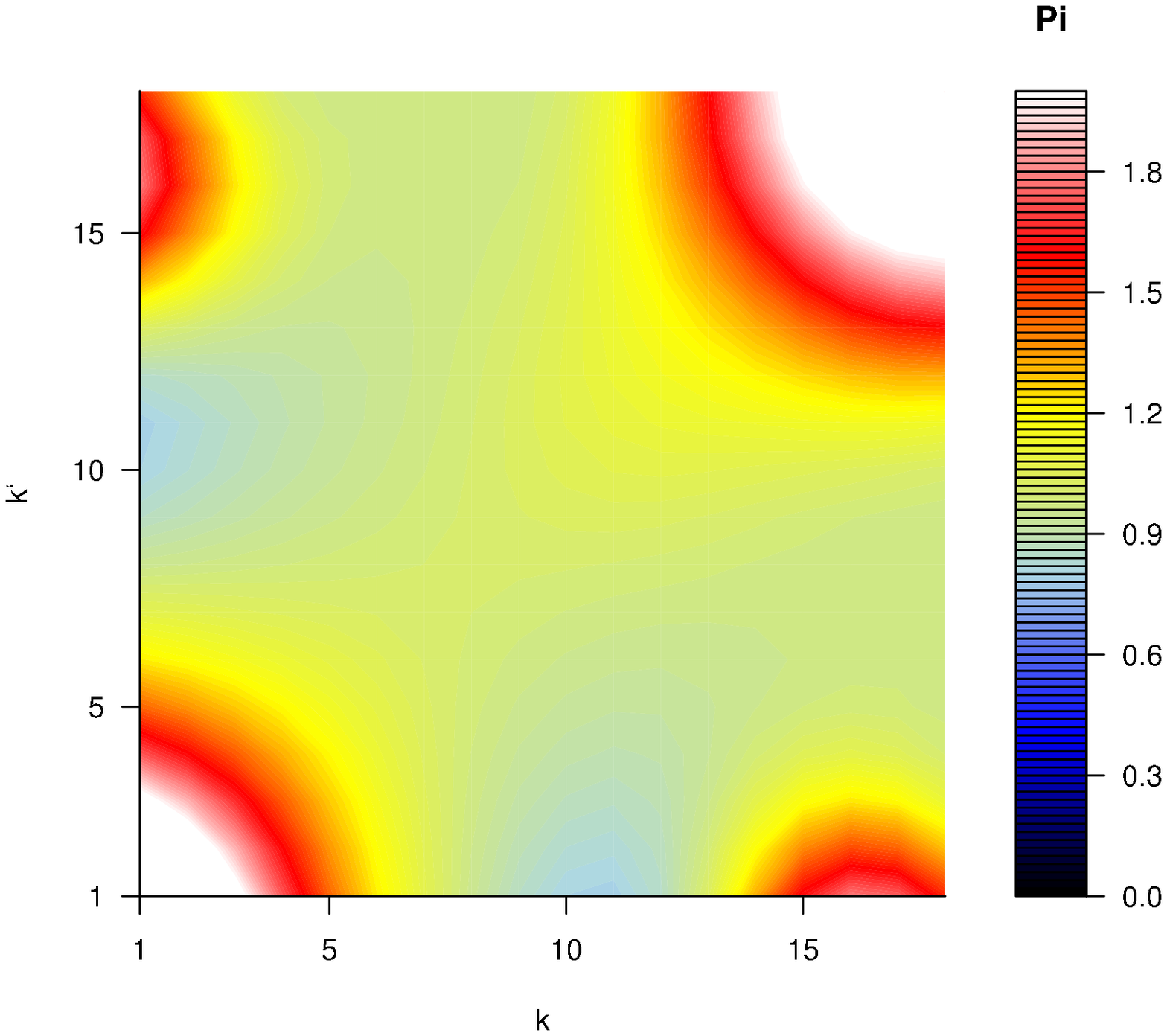}}
\put(50,200){\includegraphics[width=200\unitlength,angle=0]{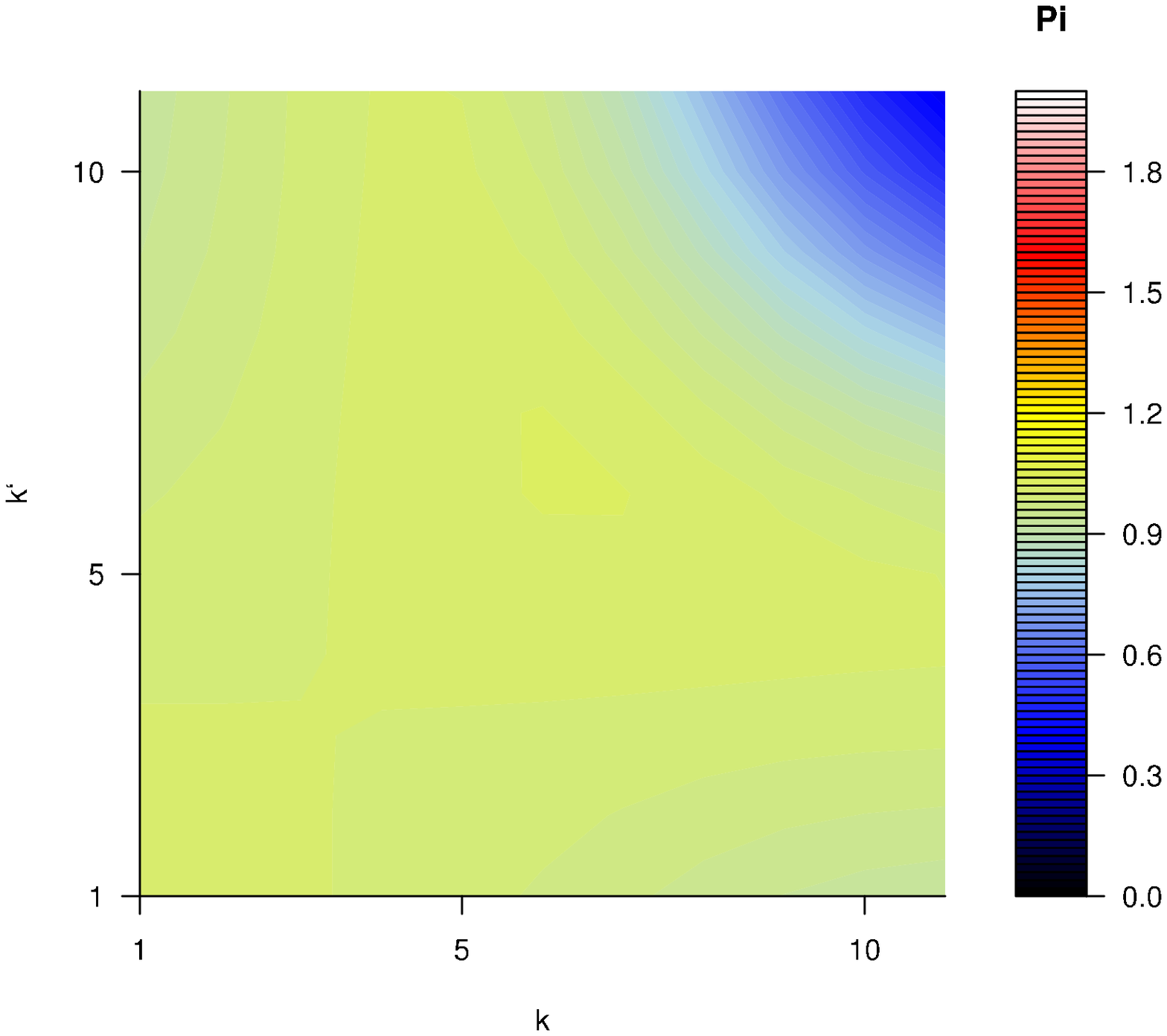}}
\put(270,200){\includegraphics[width=200\unitlength,angle=0]{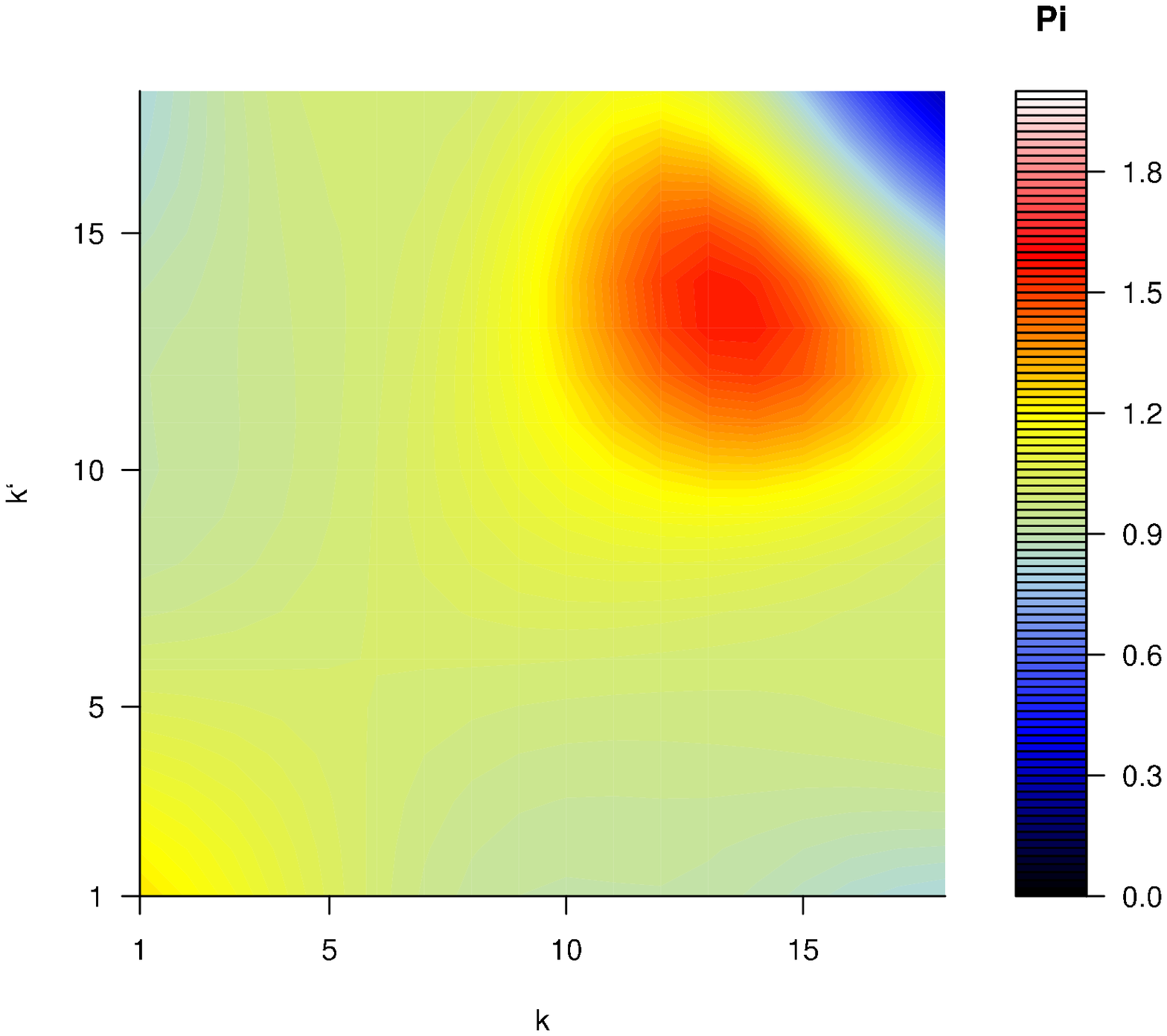}}
\put(50,10){\includegraphics[width=200\unitlength,angle=0]{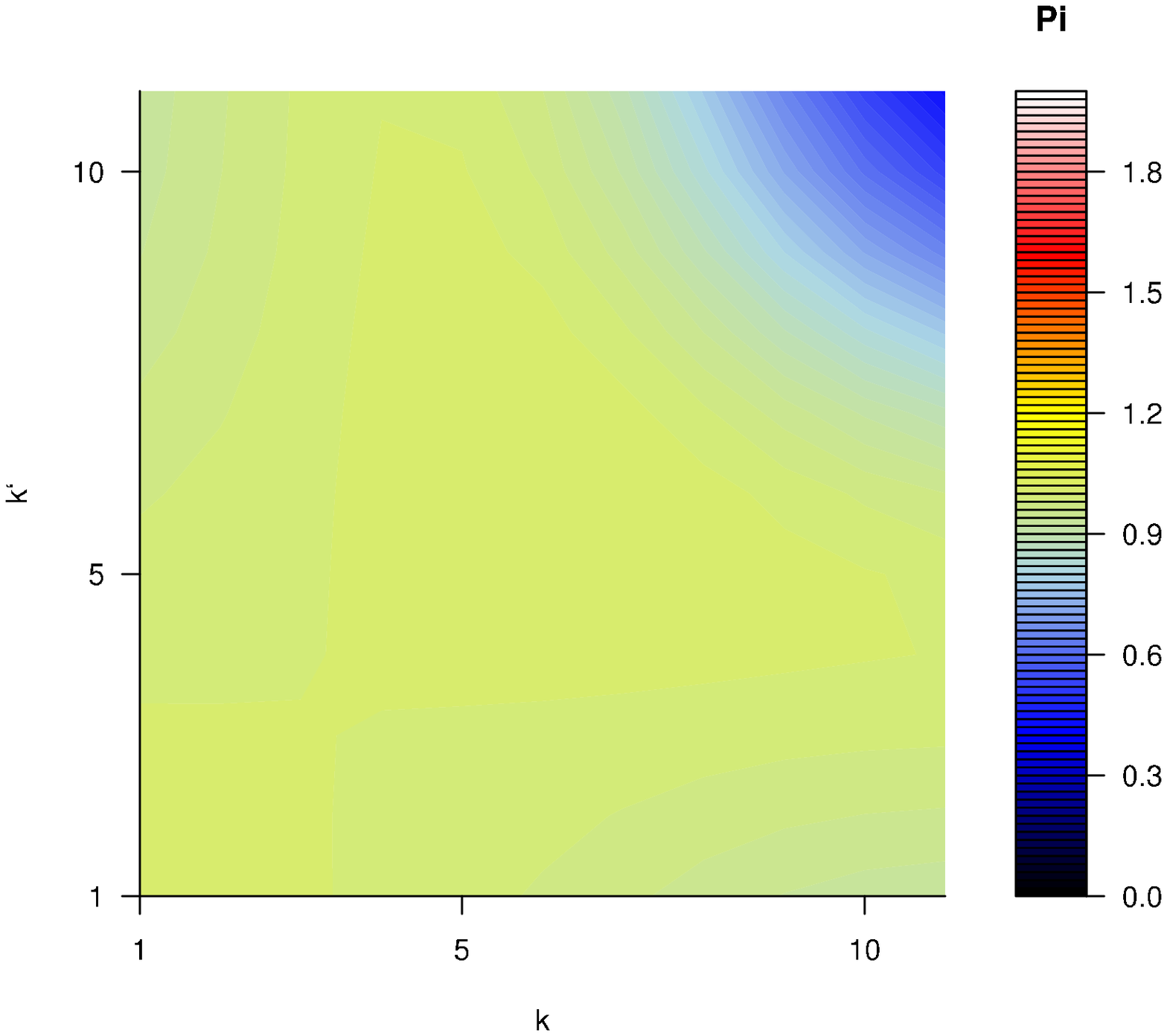}}
\put(270,10){\includegraphics[width=200\unitlength,angle=0]{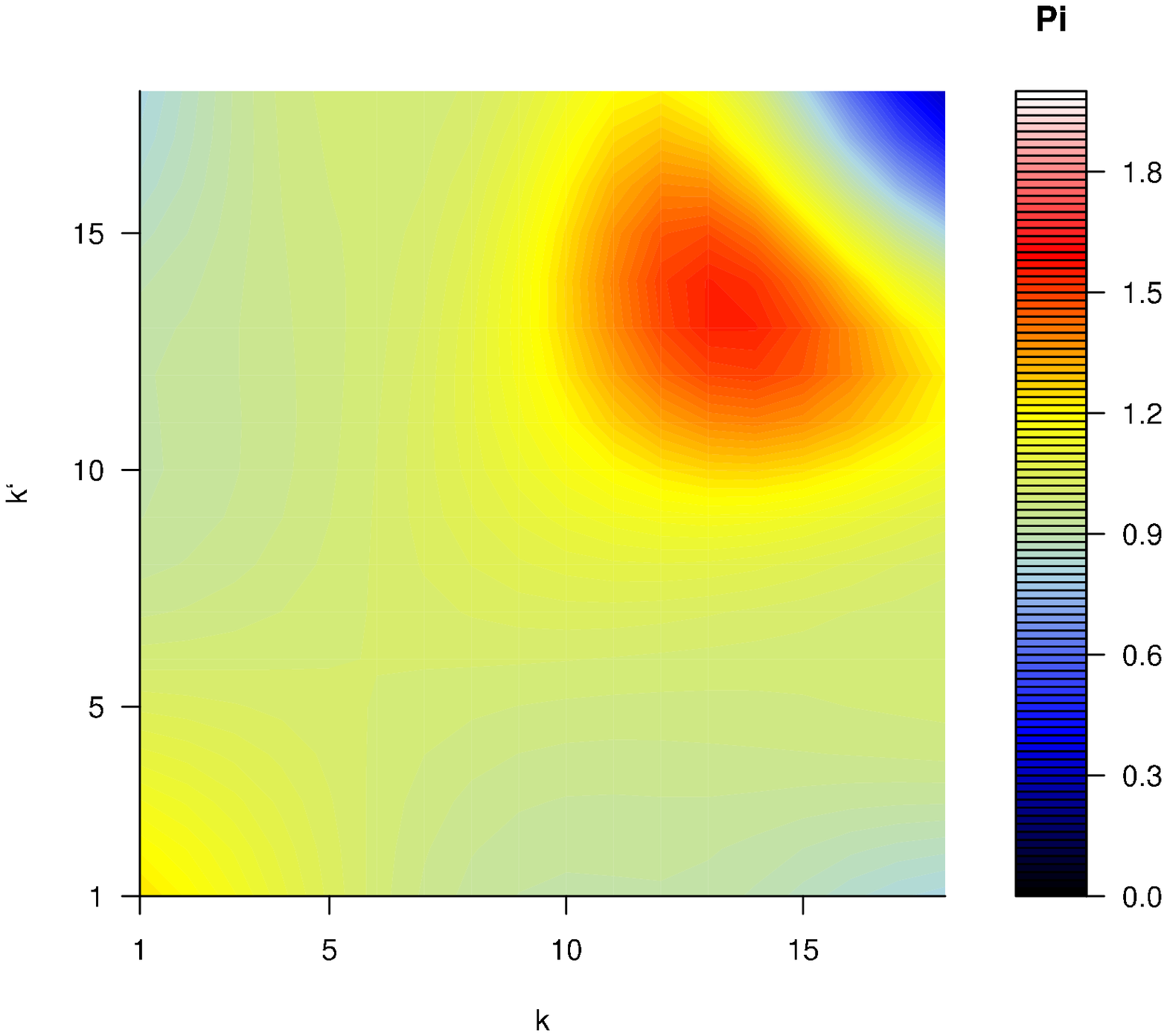}}
\end{picture}
\vspace*{-5mm}
\hspace*{-2cm}
\caption{
Normalised degree correlation function $\Pi(k,k^\prime)=W(k,k^\prime)/W(k)W(k^\prime)$
of synthetically generated Poissonian graphs with $N=3512$ and $\kav=3.72$
before (top panels) and after (middle panels)
sampling a fraction $y=0.7$ of the bonds of the original graphs (data result from averaging over $10^4$ samples)
and their respective theoretical predictions (bottom panels). 
}
\label{fig:poi_bond_undersampling}
\end{figure}
\begin{figure}[t]
 \unitlength=0.31mm
\hspace*{-13mm}
\begin{picture}(200,590)
\put(50,400){\includegraphics[width=200\unitlength,angle=0]{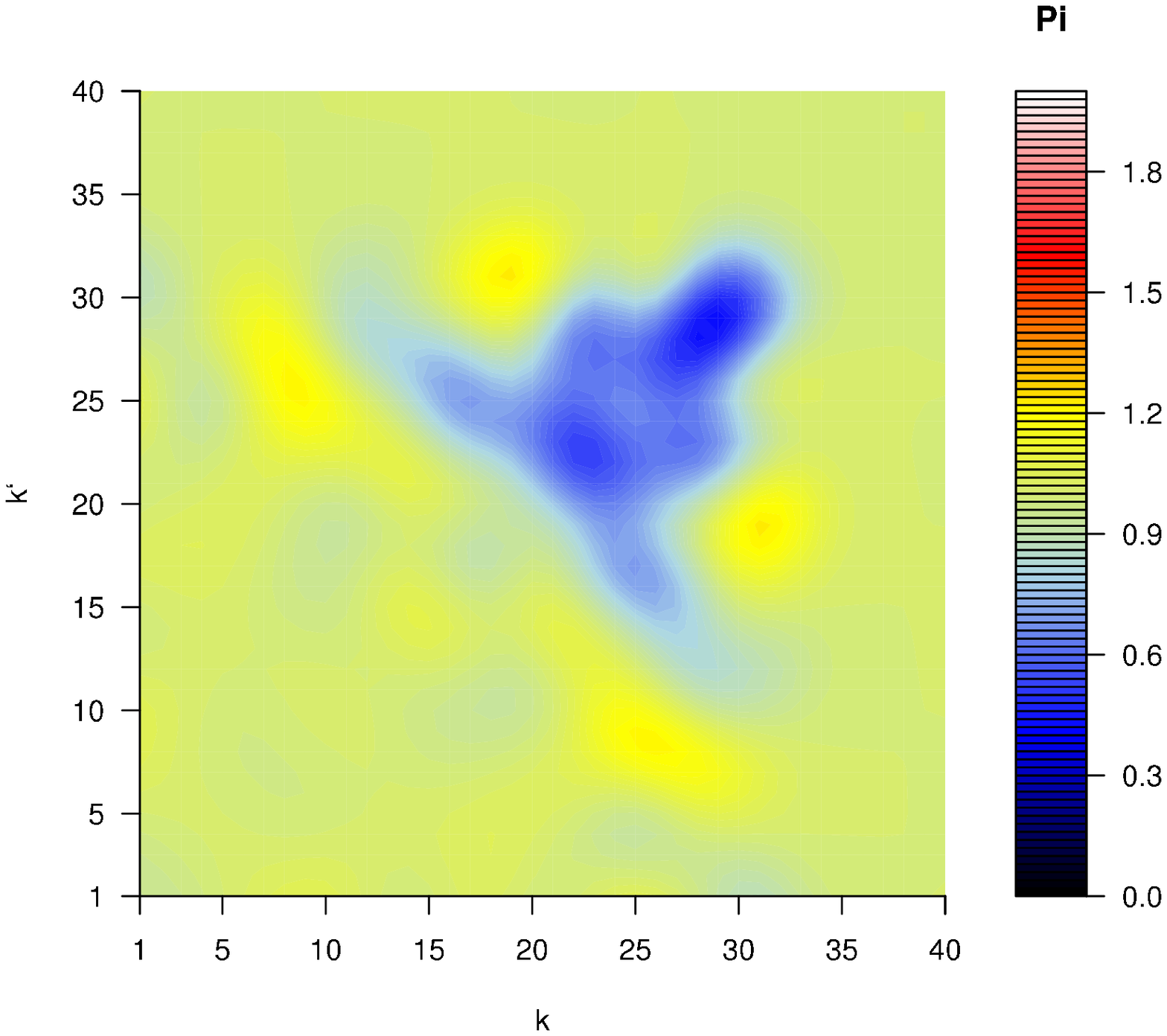}}
\put(270,400){\includegraphics[width=200\unitlength,angle=0]{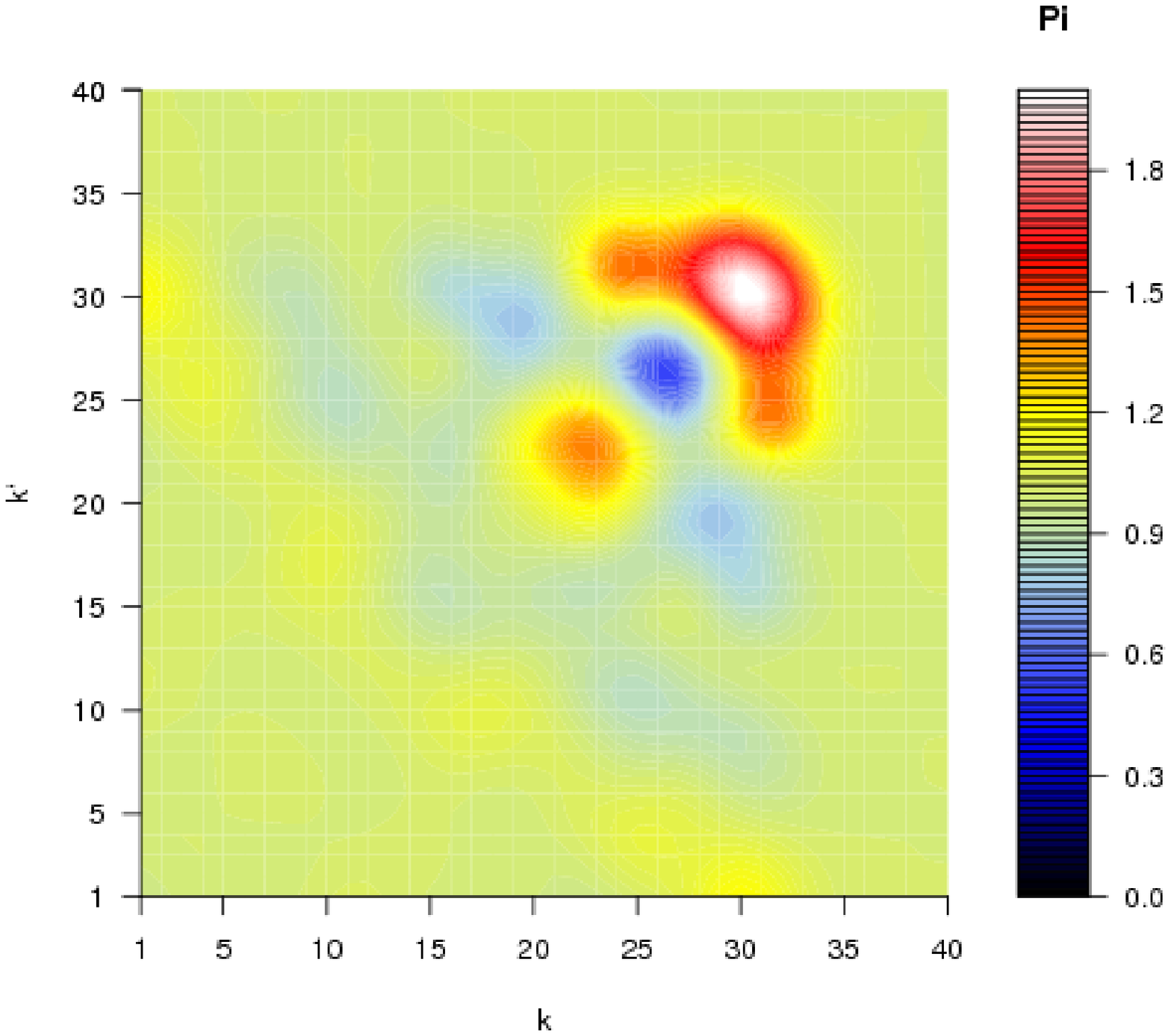}}
\put(50,200){\includegraphics[width=200\unitlength,angle=0]{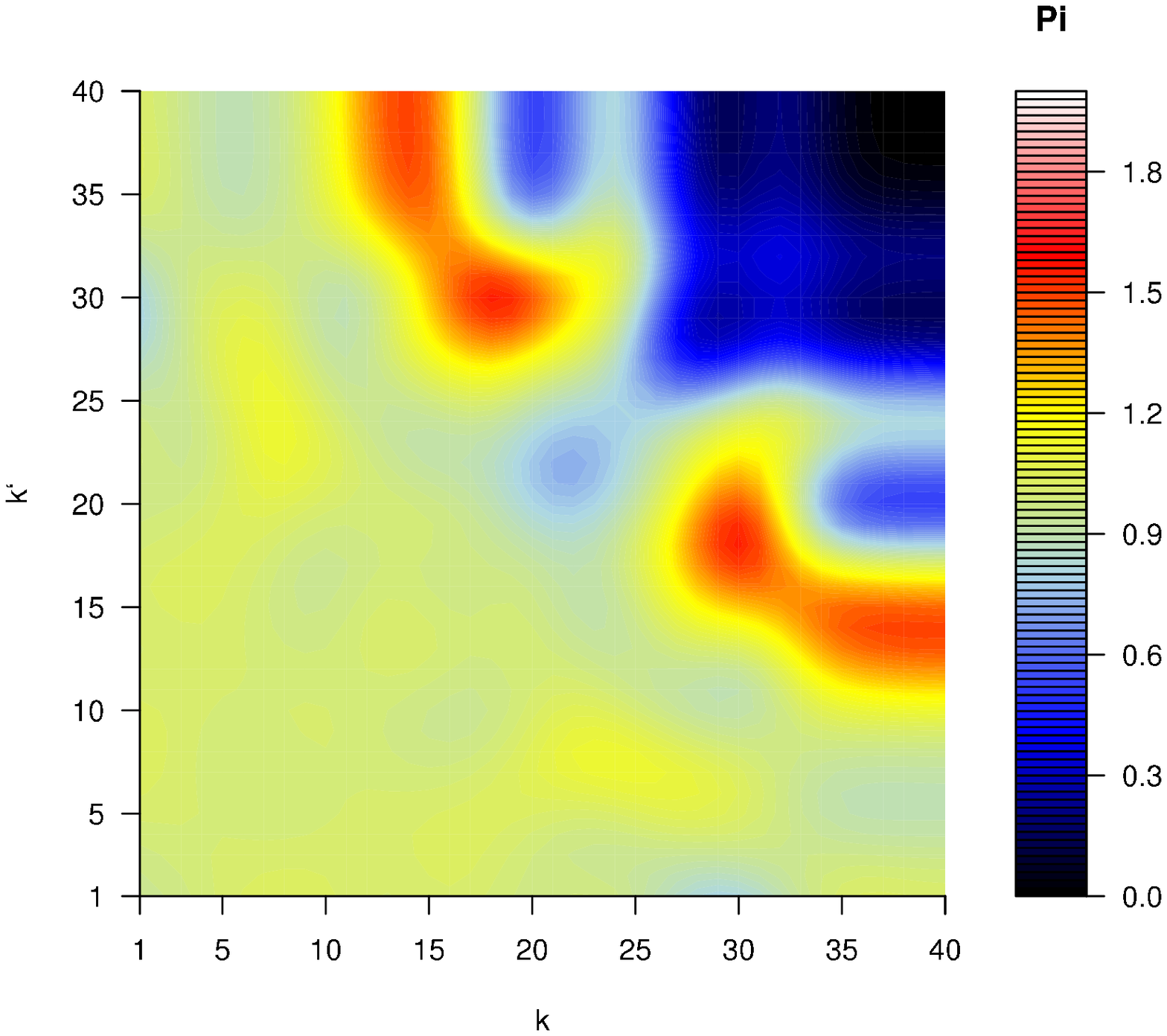}}
\put(270,200){\includegraphics[width=200\unitlength,angle=0]{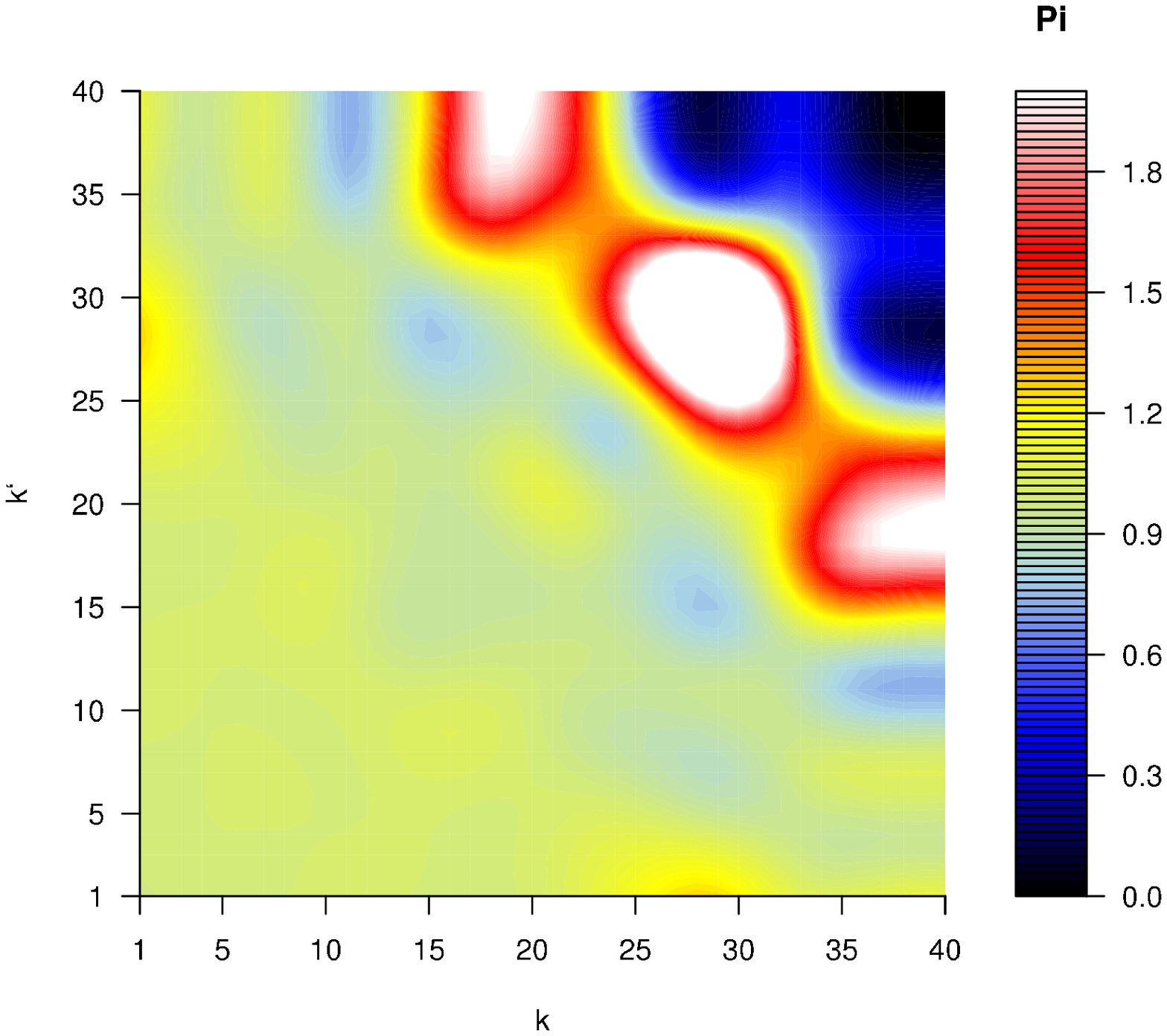}}
\put(50,10){\includegraphics[width=200\unitlength,angle=0]{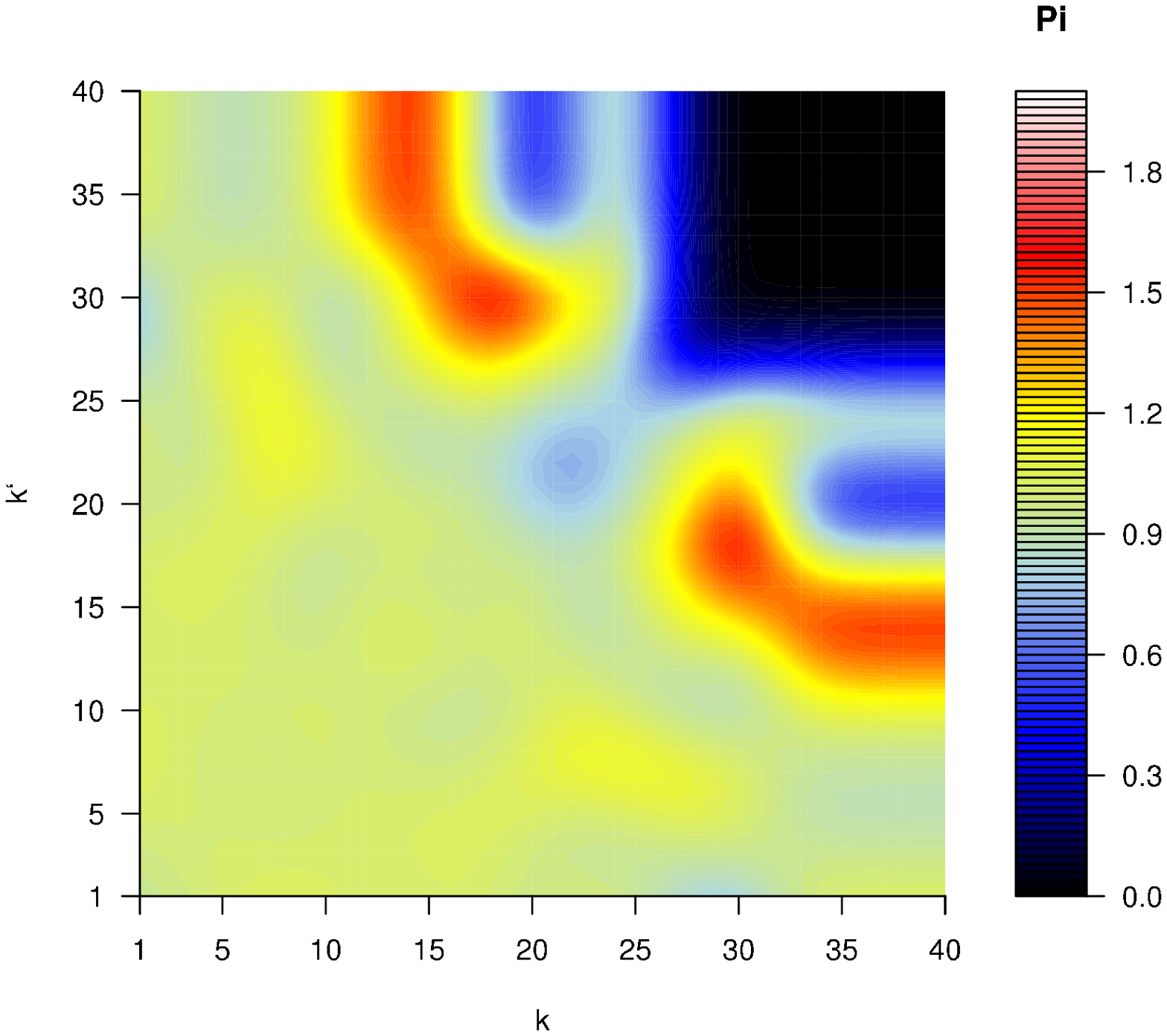}}
\put(270,10){\includegraphics[width=200\unitlength,angle=0]{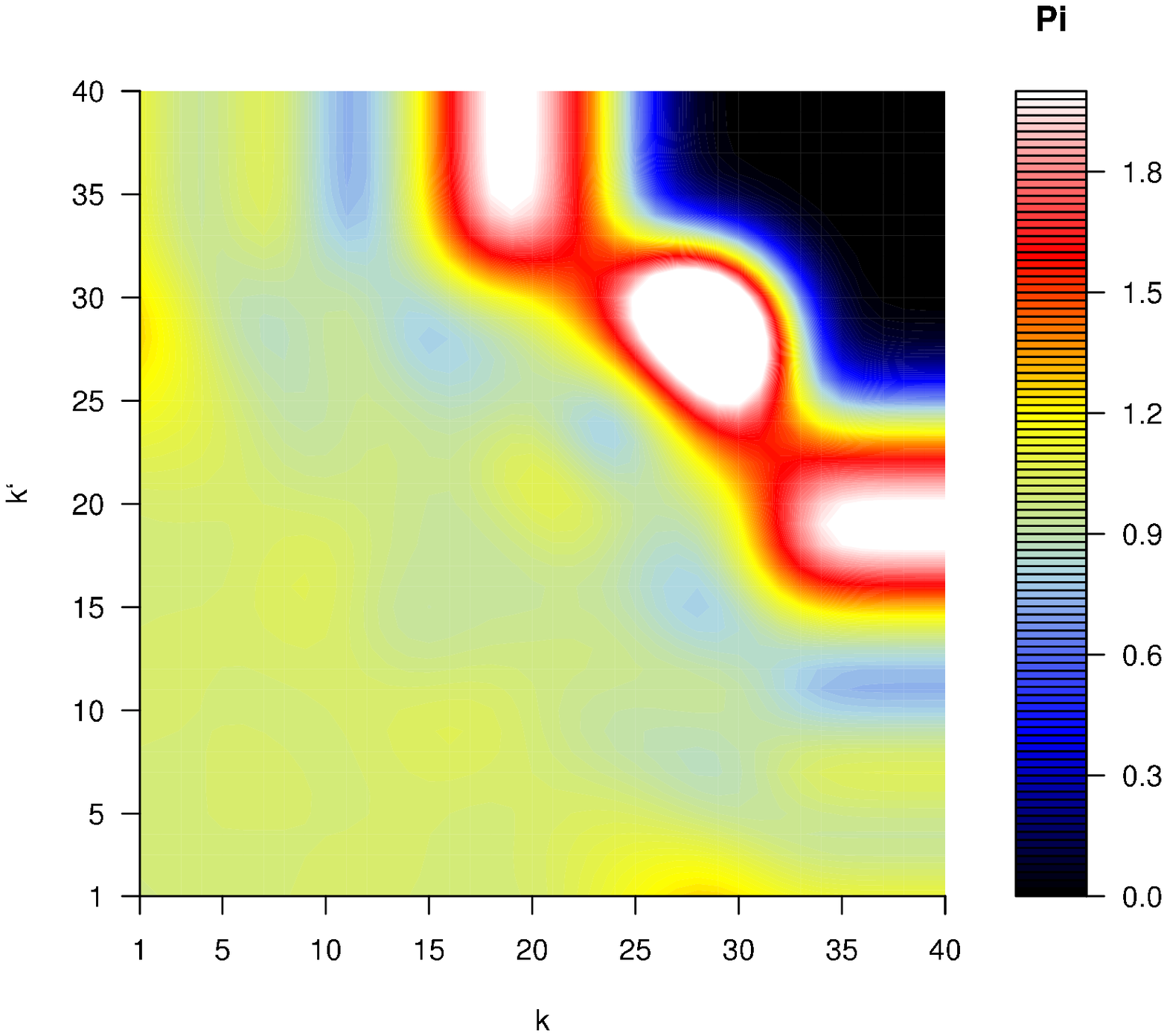}}

\end{picture}
\vspace*{-5mm}
\caption{
Normalised degree correlation function $\Pi(k,kk^\prime)$ of synthetically generated Power law
graphs with $N=3512$ and $\kav=3.72$ before (top panels) and after (middle panels)
sampling a fraction $y=0.9$ of the bonds of the original graph 
(data result from averaging over $10^4$ samples)
and 
their theoretical prediction (bottom panels).
}
\label{fig:pow_bond_undersampling}
\end{figure}
\begin{figure}[t]
 \unitlength=0.31mm
\hspace*{-13mm}
\begin{picture}(200,590)
\put(50,400){\includegraphics[width=200\unitlength,angle=0]{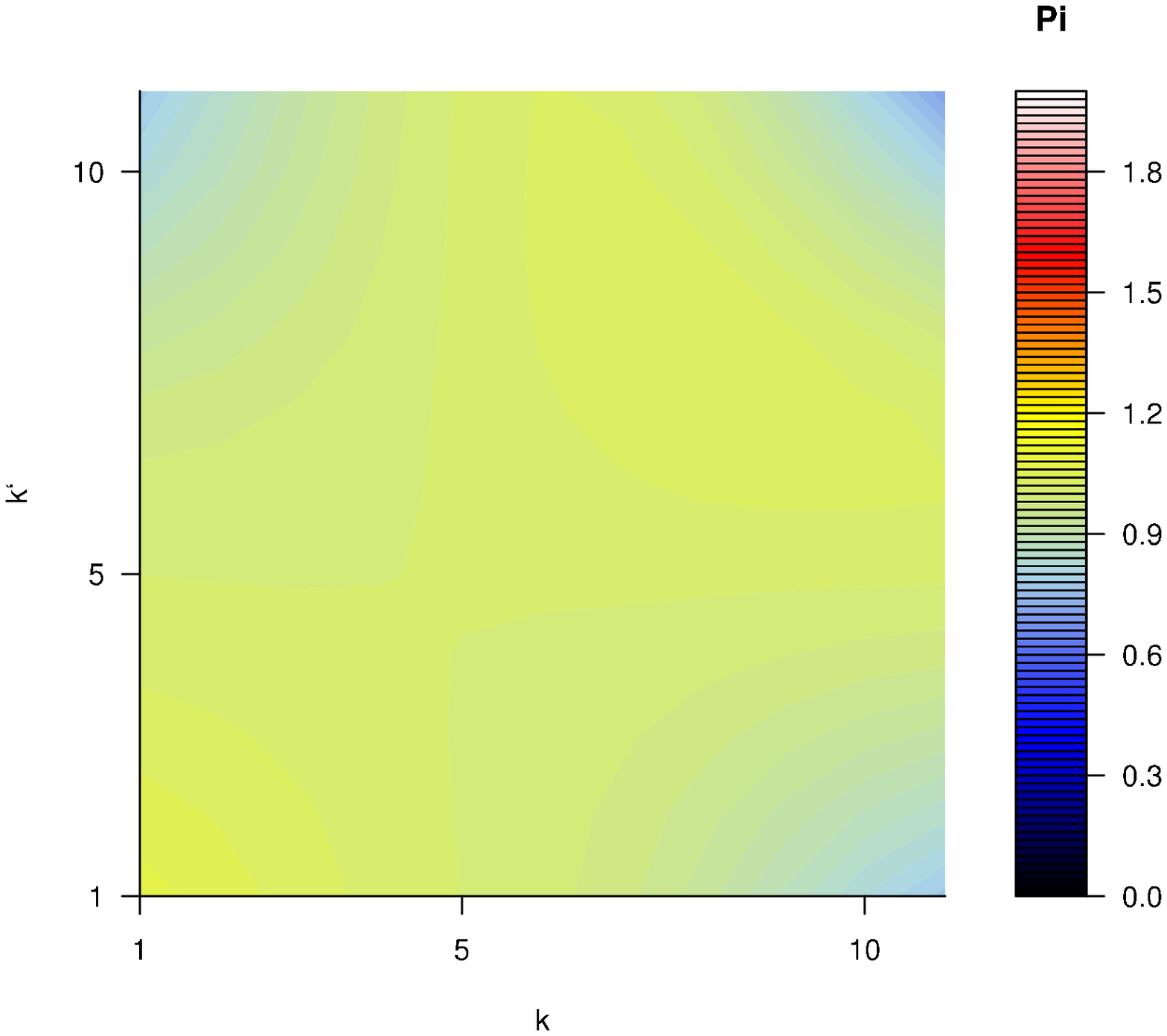}}
\put(270,400){\includegraphics[width=200\unitlength,angle=0]{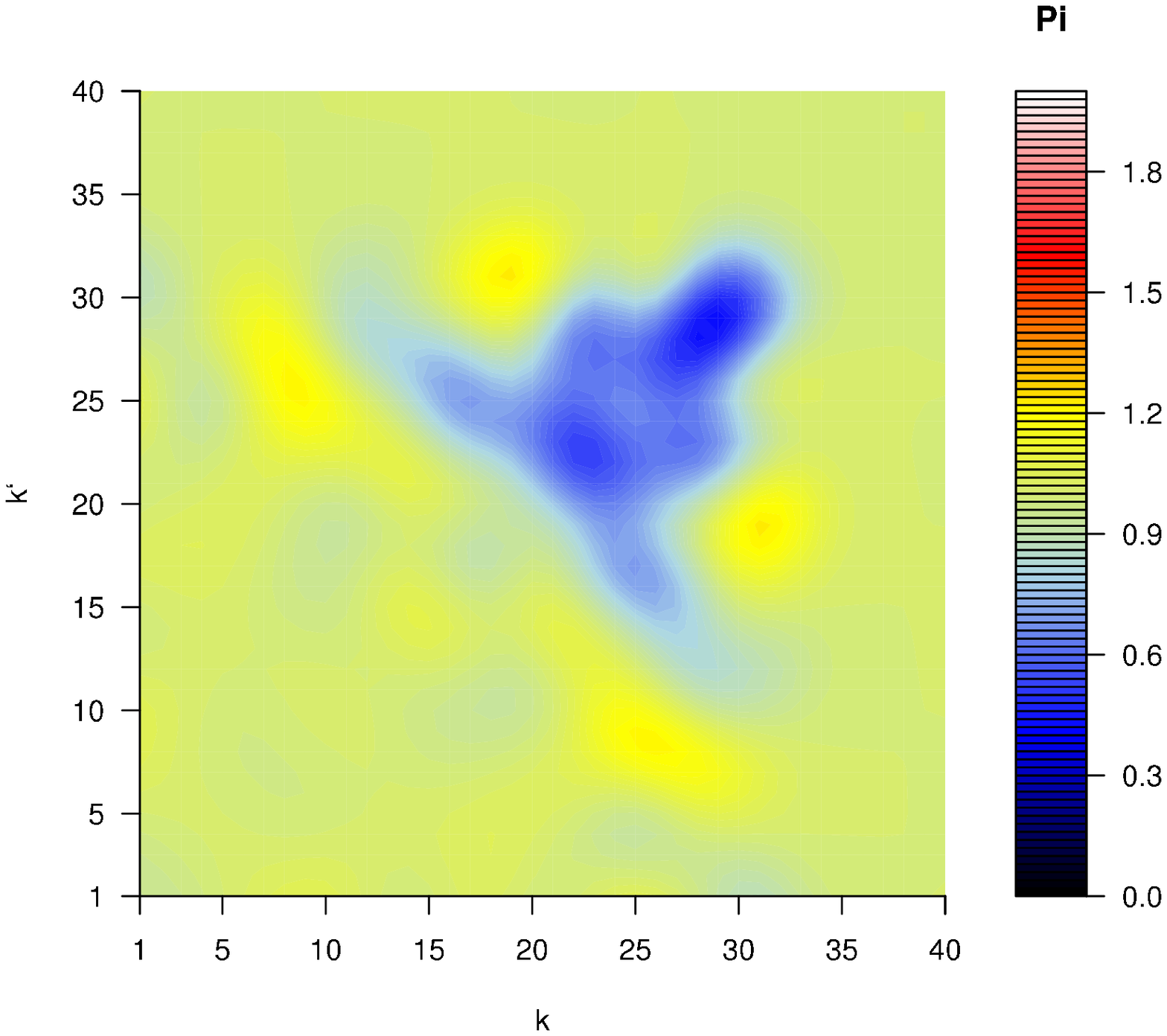}}
\put(50,200){\includegraphics[width=200\unitlength,angle=0]{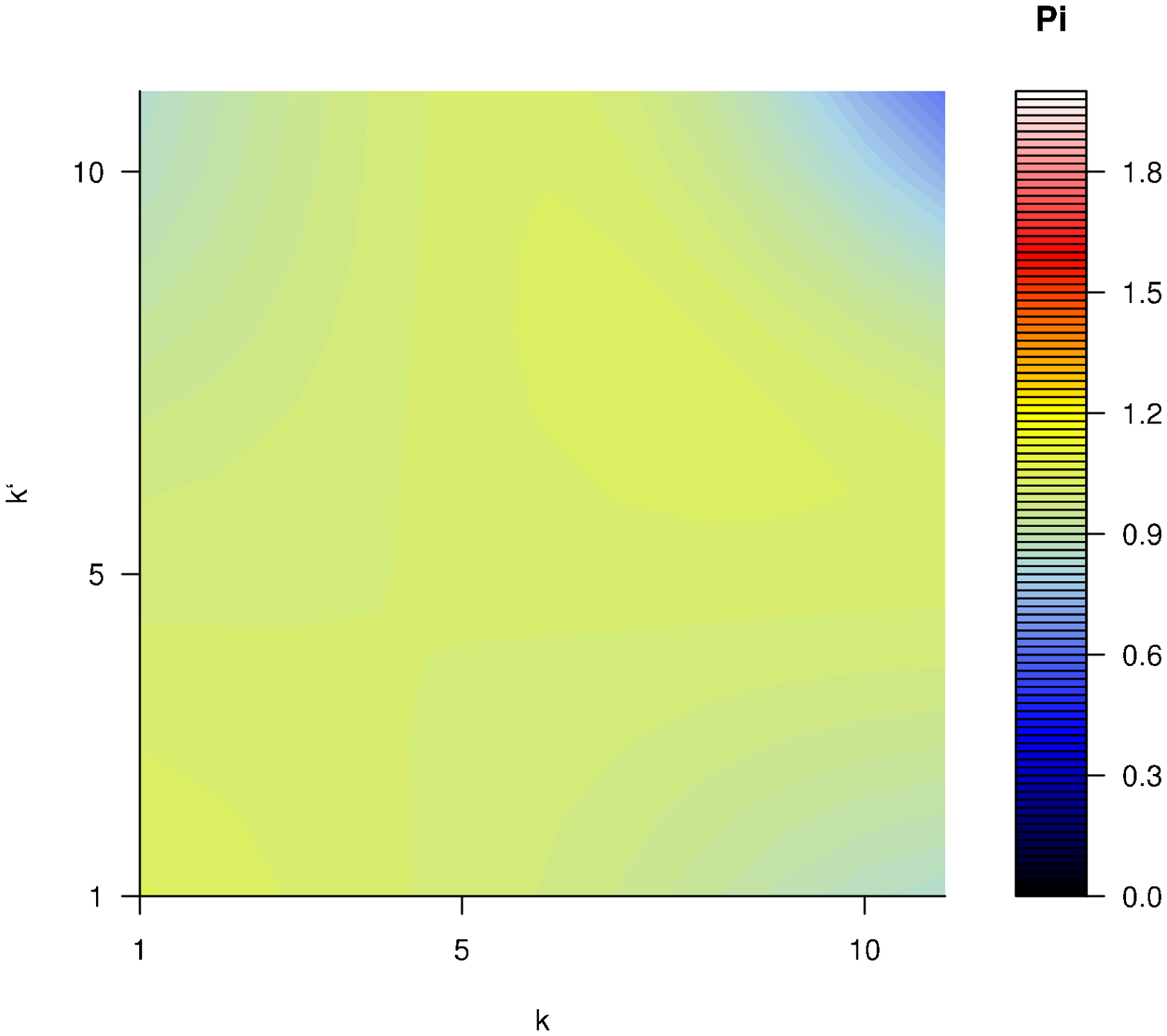}}
\put(270,200){\includegraphics[width=200\unitlength,angle=0]{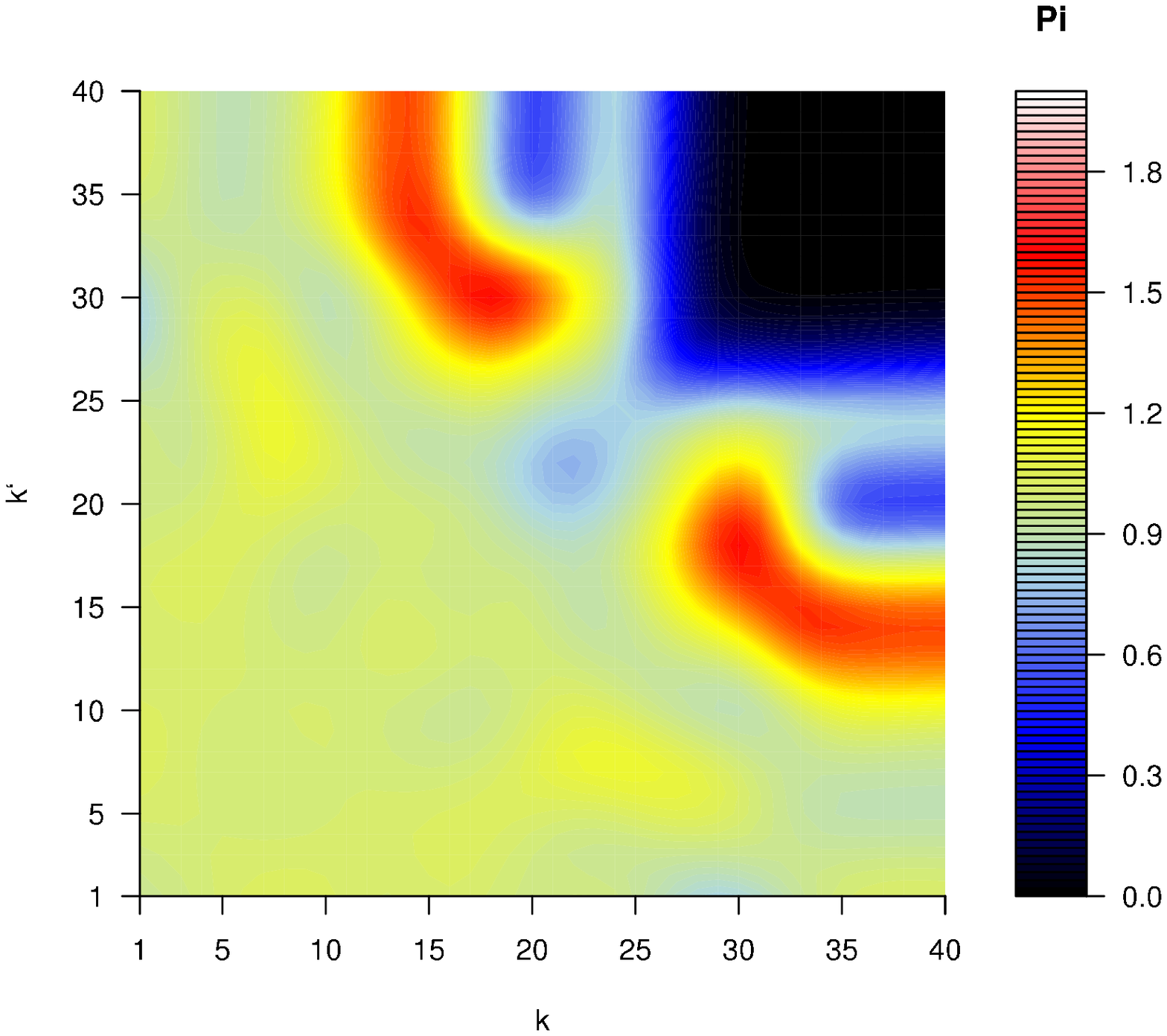}}
\put(50,10){\includegraphics[width=200\unitlength,angle=0]{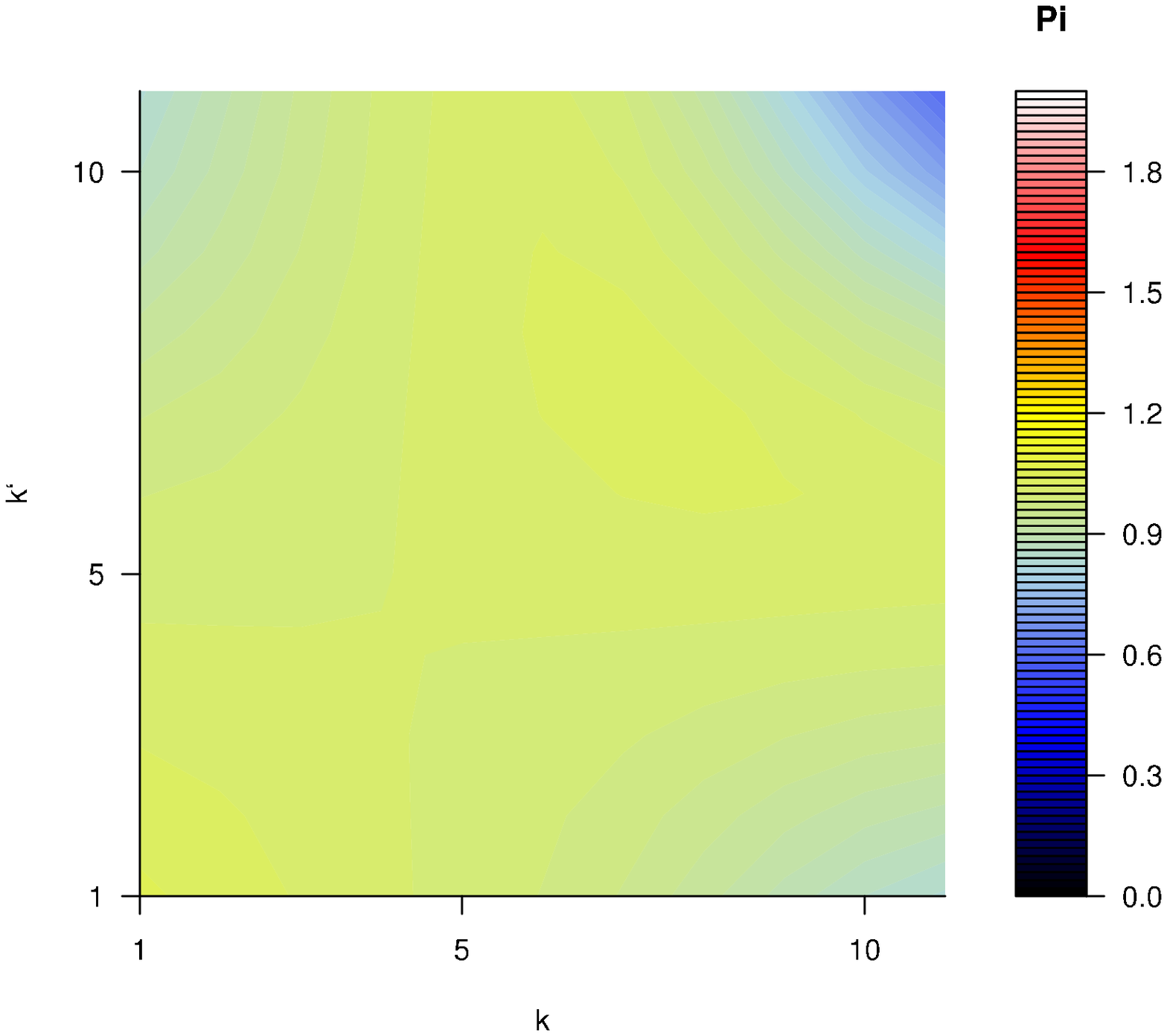}}
\put(270,10){\includegraphics[width=200\unitlength,angle=0]{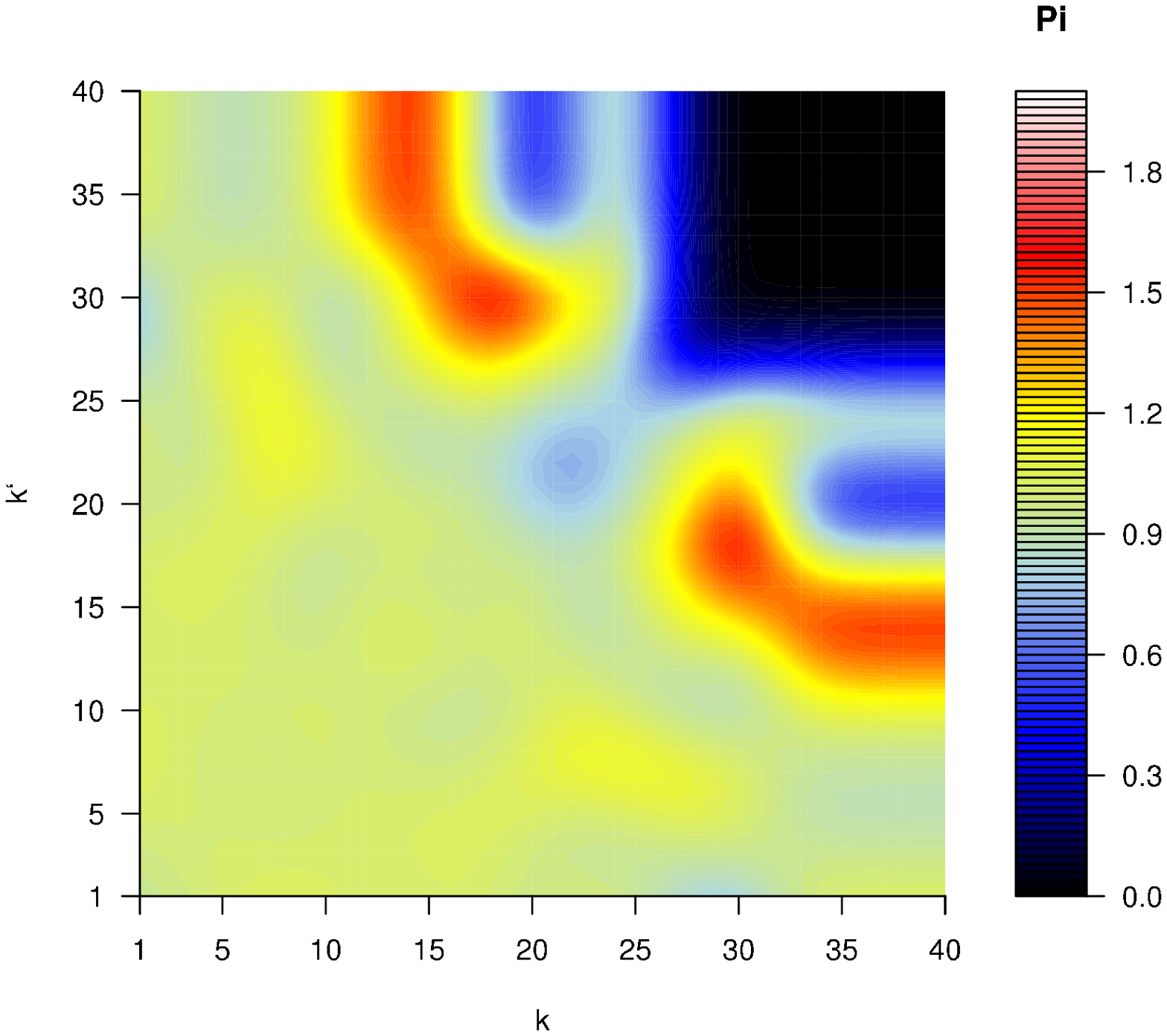}}
\end{picture}
\vspace*{-5mm}
\caption{
Normalised degree correlation function $\Pi(k,k^\prime)$ of synthetically generated Piossonian (left) and Power law
(right) graphs with $N=3512$ and $\kav=3.72$
before (top panels) and after (middle panels)
sampling a fraction $x=0.9$ of the nodes of the original graph 
(data result from averaging over $10^4$ samples)
and their theoretical prediction (bottom panels).
}
\label{fig:poi_pow_node}
\end{figure}
In order to observe how sampling protocols affect degree correlations we will 
monitor, instead of $W(k,k^\prime)$ itself, the normalised kernel $\Pi(k,k^\prime)=W(k,k^\prime)/W(k)W(k^\prime)$ which will by definition equal 
unity in the absence of degree correlations. 
Any deviation from $\Pi(k,k^\prime)=1$ will thus signal the presence of degree correlations. 
We show the predicted degree correlations in the case of unbiased bond undersampling, 
together with the corresponding results of numerical simulations, for Poissonian and Power law graphs, in figures 
\ref{fig:poi_bond_undersampling} and \ref{fig:pow_bond_undersampling} respectively.
In figure \ref{fig:poi_pow_node} we show numerical results and theoretical predictions for unbiased node undersampling from Poissonian and power law graphs. The agreement between theory and experiment is very satisfactory; deviations are small and consistent with finite size effects. 


\item {\it Unbiased bond oversampling}, i.e. $x=y=1$.
\\[2mm]
Here we have 
\be
J(k|q)=e^{-z}\frac{z^{k-1-q}}{(k-1-q)!} \,I(k\geq q+1)
\ee
so using our earlier result from equation (\ref{eq:p11z})
\be
p(k|1,1,z)=e^{-z}\sum_{q=0}^k p(q) \frac{e^{k-q}}{(k-q)!}
\ee
we may write 
\be
\sum_{q\geq 0} p(q)J(k|q)=p(k-1|1,1,z)
\label{eq:sumpJ}
\ee
which leads to the transparent expression
\bea
\hspace*{-3cm}W(k,k^\prime|1,\!1,\!z)&=&
\frac{z}{\kav+z} \left[
p(k\!-\!1|1,\!1,\!z)p(k^\prime\!-\!1|1,\!1,\!z)
+
e^{-2z}\frac{\kav}{z}\sum_{q,q^\prime=1}^{k,k^\prime}
W(q,q^\prime)\frac{z^{k-q}}{(k-q)!}\frac{z^{k^\prime-q^\prime}}{(k^\prime-q^\prime)!}\right]
\label{eq:W_unbiased_over}\nonumber
\\
\eea
%
We note for later that substituting (\ref{eq:sumpJ}) into (\ref{eq:general_p}) and 
bearing in mind that ${\cal L}(k|q)={\cal J}(k|q-1)$, $a(q)=z$ and $b(q)=1$,
we have
\be
p(k|1,1,z)=\frac{1}{k}\Big [z\,p(k-1|1,1,z)+\sum_{q\geq 1} p(q)q J(k|q-1)\Big]
\ee
which yields
\be
p(k-1|1,1,z)=\frac{k}{z}p(k|1,1,z)-\frac{\kav}{z}\sum_{q\geq 1} W(q) J(k|q-1)
\ee
where $W(k)=k p(k)/\kav$.

We now study the effects of oversampling on graphs without degree 
correlations. 
Denoting
\be
\sk=\sum_{q \geq 1} W(q)J(k|q-1)
\ee
which is $z$-dependent via the function $J$, 
we may rewrite (\ref{eq:W_unbiased_over}) as
\bea
W(k,k'|1,\!1,\!z)&=&
\frac{z}{\kav+z} \left[
\left(\frac{k}{z}p(k|1,\!1,\!z)-\frac{\kav}{z}\sk\right)
\left(\frac{k'}{z}p(k'|1,\!1,\!z)-\frac{\kav}{z}\skp\right)\right.
\nonumber\\
&&+\left.
\frac{\kav}{z}\sum_{q,q'\geq 1} W(q,q') J(k|q-1)J(k^\prime|q^\prime-1) \right]
\eea
If the original graph has no degree correlation, i.e.
\be
W(q,q')=W(q)W(q')=p(q)p(q')\frac{qq'}{\kav^2}
\ee
the sampled graph will have degree correlation
\bea
\hspace*{-2cm}
W(k,k'|1,\!1,\!z)&=&\frac{z}{\kav+z} \left[
\left(\frac{k}{z}p(k|1,\!1,\!z)-\frac{\kav}{z}\sk\right)
\left(\frac{k'}{z}p(k'|1,\!1,\!z)-\frac{\kav}{z}\skp\right)+
\frac{\kav}{z}\sk\skp\right]
\nonumber
\\
&&\hspace*{-3cm}=\frac{\kav}{z}\left[\frac{\kav+z}{\kav}W(k|1,\!1,\!z)W(k'|1,\!1,\!z)+
\sk\skp-W(k|1,\!1,\!z)\skp-W(k'|1,\!1,\!z)\sk
\right]
\nonumber
\\
&&\hspace*{-3cm}=\frac{\kav}{z}\left[(W(k|1,\!1,\!z)-\sk)(W(k'|1,\!1,\!z)-\skp)+\frac{z}{\kav}W(k|1,\!1,\!z)W(k'|1,\!1,\!z)
\right]
\nonumber
\\
&&\hspace*{-3cm}=W(k|1,\!1,\!z)W(k'|1,\!1,\!z)+\frac{\kav}{z}(W(k|1,\!1,\!z)-\sk)(W(k'|1,\!1,\!z)-\skp)
\label{eq:induced_correlation}
\eea
where we have used $W(k,k^\prime|1,1,z)=k\, p(k|1,1,z)/(\kav+z)$, in accordance with (\ref{eq:connection}) and (\ref{eq:unbiased_kav}).
For $z=0$, $J(k|q)=\delta_{k,q+1}$ and $W(k|0)=S_0(k)=W(k)$ so the second term in (\ref{eq:induced_correlation}) 
vanishes, however for $z\neq 0$ this will be generally different from zero:
{\it crucially}, but not unexpectedly, 
oversampling from a graph without degree correlations
automatically introduces degree correlations. 
\end{itemize}

\begin{figure}[t]
 \unitlength=0.31mm
\hspace*{-13mm}
\begin{picture}(200,590)
\put(50,400){\includegraphics[width=190\unitlength,angle=0]{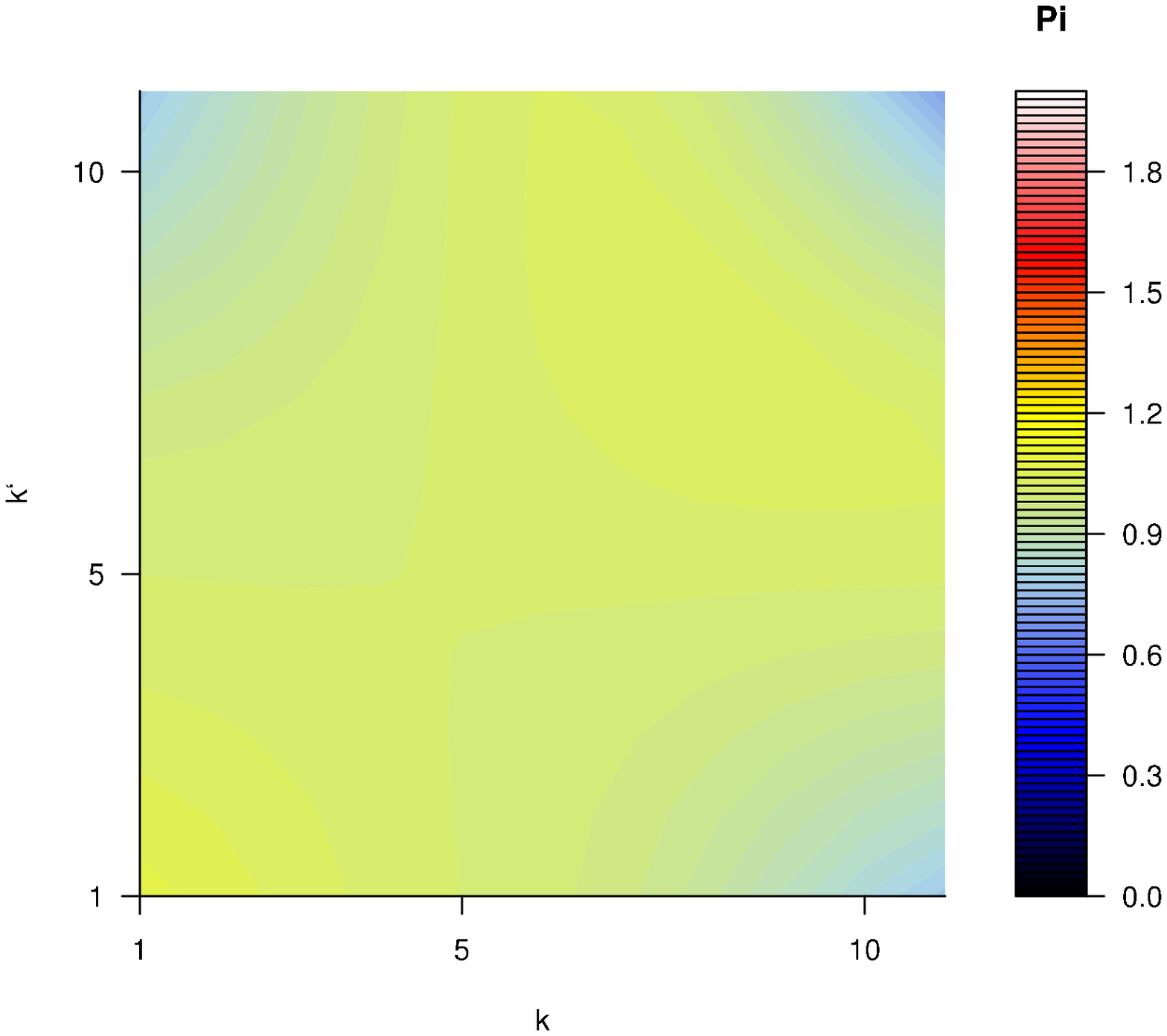}}
\put(270,400){\includegraphics[width=190\unitlength,angle=0]{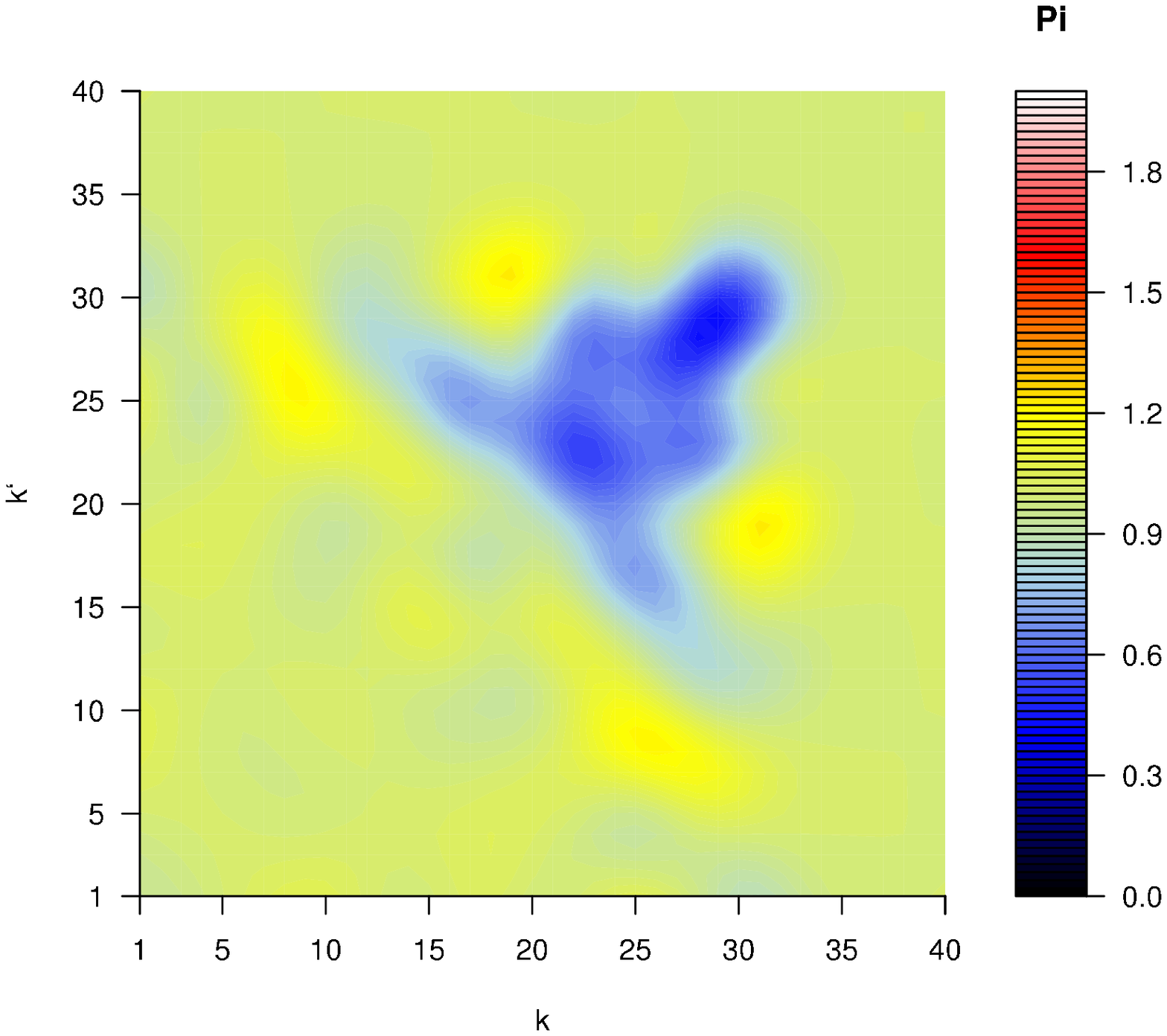}}
\put(50,200){\includegraphics[width=190\unitlength,angle=0]{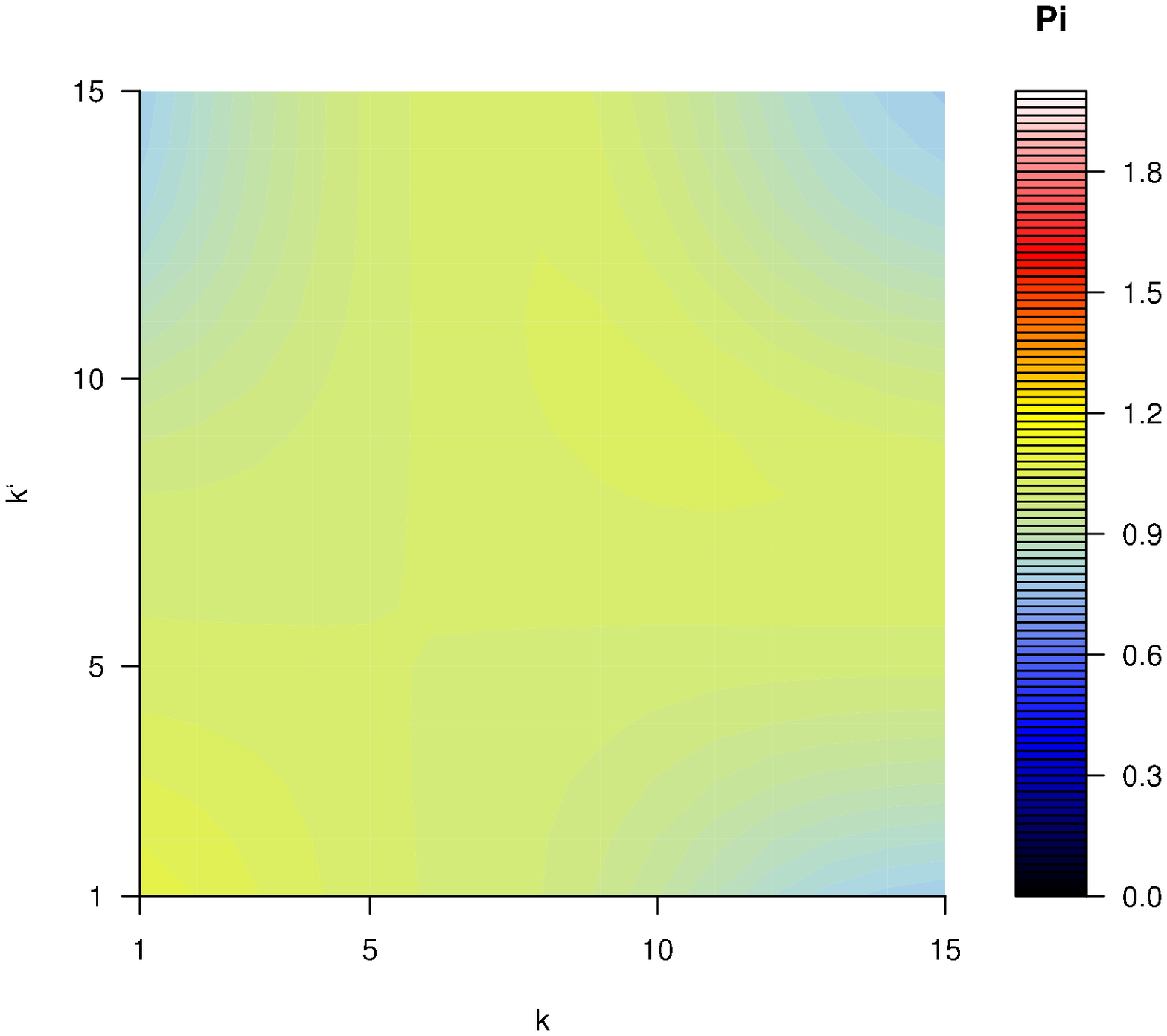}}
\put(270,200){\includegraphics[width=190\unitlength,angle=0]{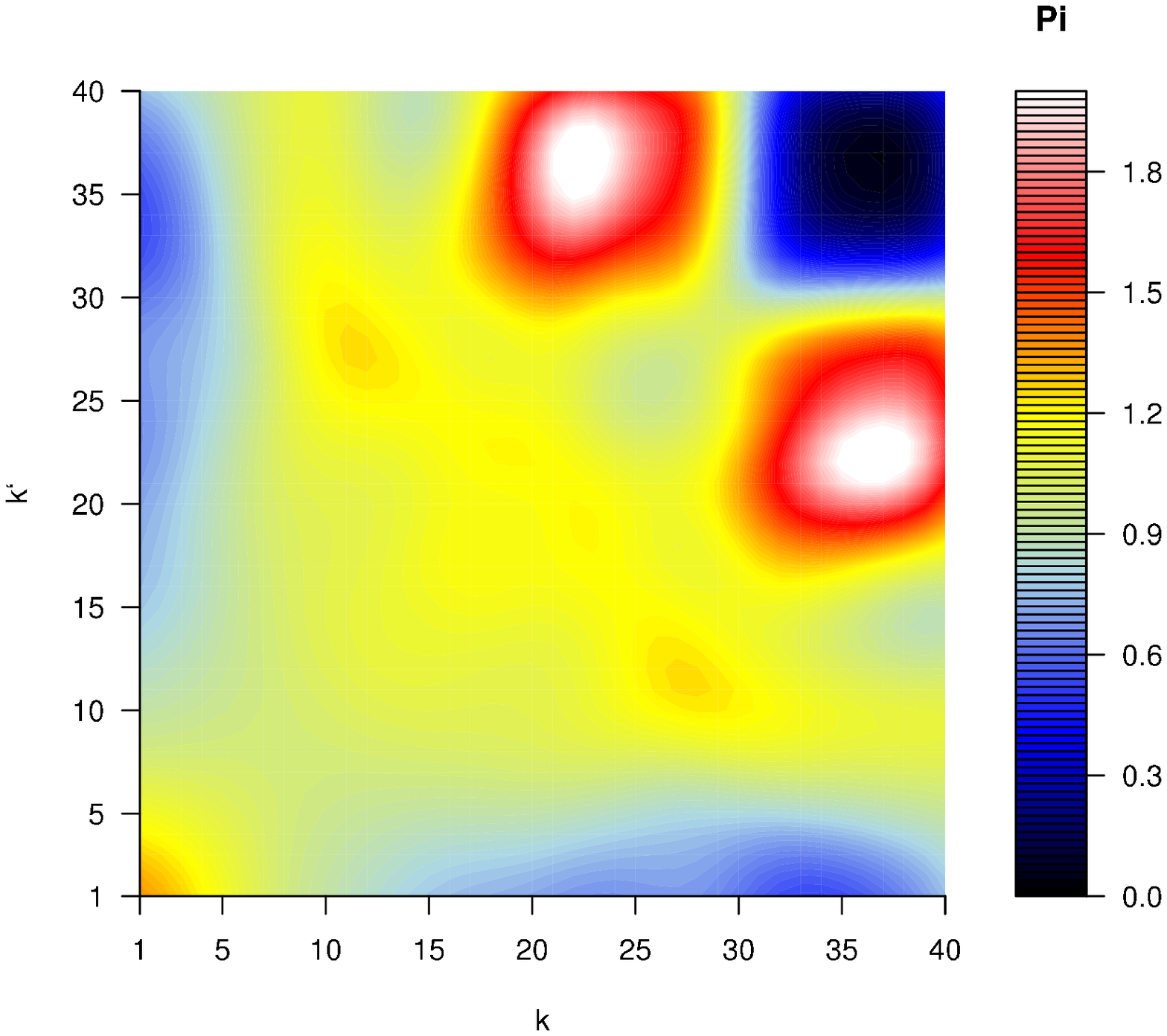}}
\put(50,10){\includegraphics[width=190\unitlength,angle=0]{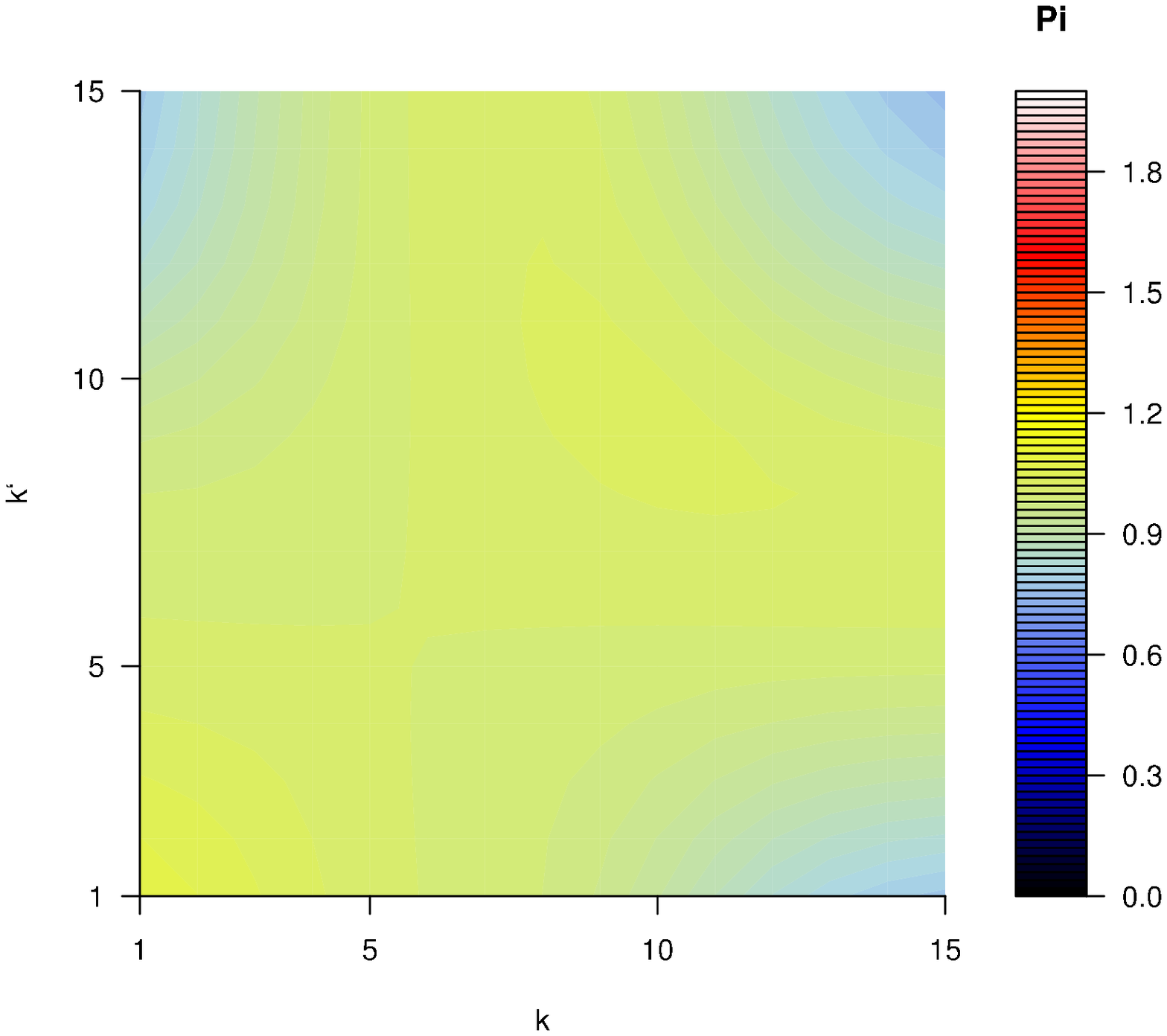}}
\put(270,10){\includegraphics[width=190\unitlength,angle=0]{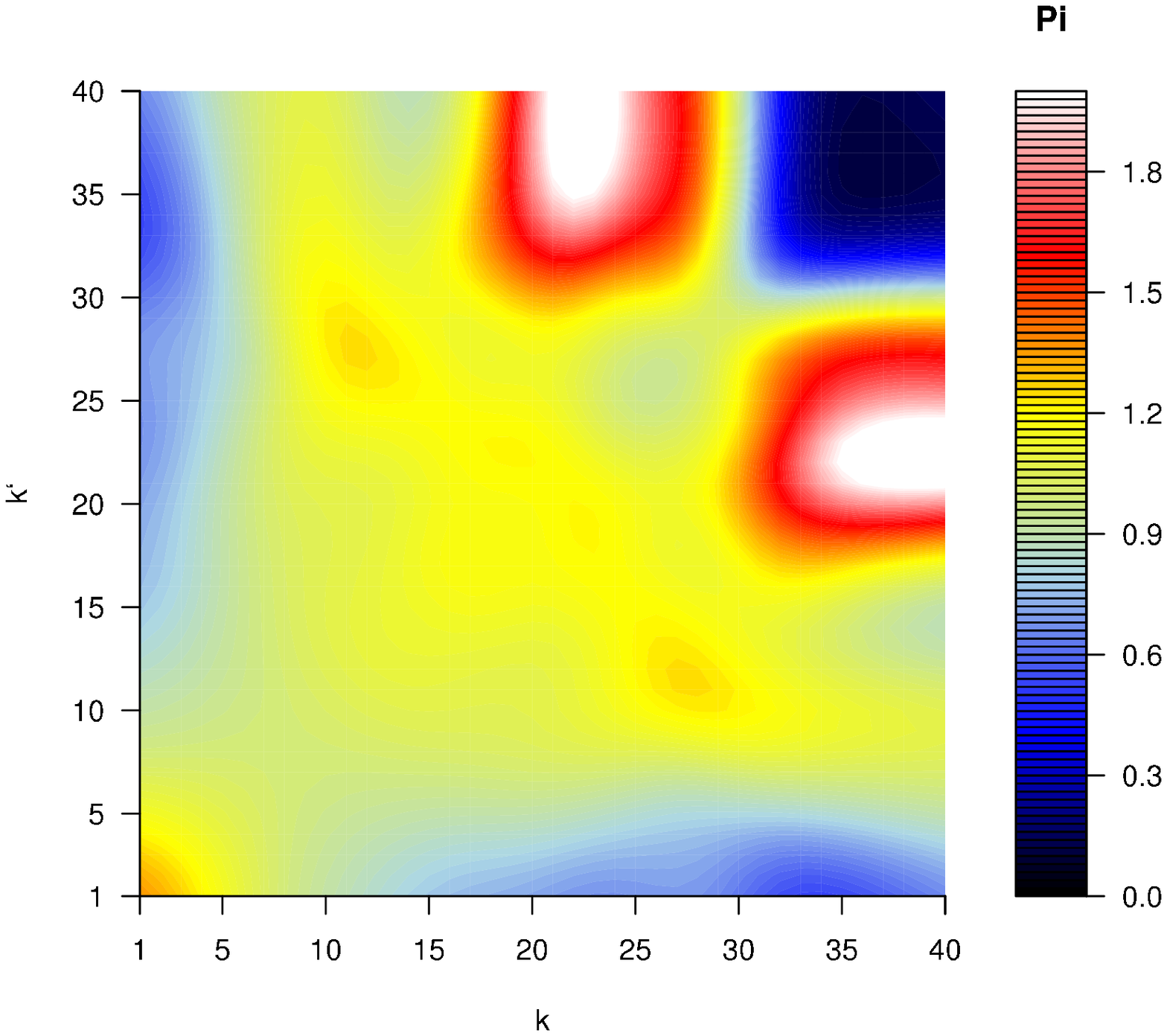}}
\end{picture}
\vspace*{-5mm}
\caption{
Normalised degree correlation function $\Pi(k,k^\prime)$ of synthetically generated Poissonian (left) 
and Power law (right) graphs with $N=3512$ and $\kav=3.72$
before (top panels) and after (middle panels)
adding a fraction $z/N$ of bonds, with 
$z=1$ (left) and $z=2$ (right), and
their respective theoretical predictions (bottom panels). Data obtained by averaging over $10^4$ samples.
}
\end{figure}

\subsection{Degree correlations for biased sampling}
Let us now work out (\ref{eq:general_W}) for the types of biased sampling considered above.

\begin{itemize}
\item {\it Biased node undersampling}, i.e. $x(k)=\alpha k/k_{\rm max}$, $y(k,k')=1$, $z(k,k')=0$
\\
Here we have
\begin{eqnarray}
&&
\hspace*{-2cm}
b(q)= \frac{\alpha \kav}{k_{\rm max}q p(q)} \sum_{q^\prime} W(q,q^\prime) q^\prime,~~~~~~~~~ a(q)=0
\\
&&
\hspace*{-2cm}
{\cal L}(k|q)=
\left(
\begin{array}{l}
q-1\\
k-1
\end{array}
\right)
\left(
\frac{\alpha \kav}{k_{\rm max}q p(q)}\sum_{q''}q''W(q,q'')
\right)^{k-1}
\left(
1-\frac{\alpha \kav}{k_{\rm max}q p(q)}\sum_{q''} q''W(q,q'')
\right)^{q-k}
\end{eqnarray}
so our equations reduce to
\bea
&&
\hspace*{-2cm}W(k,k'|x)=\frac{\alpha \kav^2}{k_{\rm max}\bra k^{(3)}\ket}\sum_{qq'}W(q,q')qq'
\nonumber\\
&&\hspace*{-2cm}
\times\left(
\begin{array}{l}
q-1\\
k-1
\end{array}
\right)
\left(
\frac{\alpha \kav}{k_{\rm max}q p(q)}\sum_{q''}q''W(q,q'')
\right)^{k-1}
\left(
1-\frac{\alpha \kav}{k_{\rm max}q p(q)}\sum_{q''}q''W(q,q'')
\right)^{q-k}
\nonumber\\
&&\hspace*{-2cm}\times
\left(
\begin{array}{l}
q'-1\\
k'-1
\end{array}
\right)
\left(
\frac{\alpha \kav}{k_{\rm max}qp(q)}\sum_{q''} q''W(q',q'')
\right)^{k'-1}
\left(
1-\frac{\alpha \kav}{k_{\rm max}q' p(q')}\sum_{q''} q''W(q',q'')
\right)^{q'-k'}
\eea
\item {\it Biased bond undersampling}, i.e. $x(k)=1$, $y(k,k^\prime)=\alpha k k^\prime/k^2_{\rm max}$, $z(k,k')=0$
\\
For this choice we obtain
\begin{eqnarray}
&&
\hspace*{-2cm}
b(q)= \frac{\alpha \kav}{k^2_{\rm max} p(q)} \sum_{q^\prime} W(q,q^\prime) q^\prime,~~~~~~~~~ a(q)=0
\\
&&
\hspace*{-2cm}
{\cal L}(k|q)=
\left(
\begin{array}{l}
q-1\\
k-1
\end{array}
\right)
\left(
\frac{\alpha \kav}{k^2_{\rm max} p(q)}\sum_{q''}q''W(q,q'')
\right)^{k-1}
\left(
1-\frac{\alpha \kav}{k^2_{\rm max} p(q)}\sum_{q''} q''W(q,q'')
\right)^{q-k}
\end{eqnarray}
which leads to
\bea
&&
\hspace*{-1cm}W(k,k'|x)=\frac{\alpha \kav^2}{k_{\rm max}\bra k^{(3)}\ket}\sum_{qq'}W(q,q')qq'
\nonumber\\
&&\hspace*{-1cm}
\times\left(
\begin{array}{l}
q-1\\
k-1
\end{array}
\right)
\left(
\frac{\alpha \kav}{k^2_{\rm max} p(q)}\sum_{q''}q''W(q,q'')
\right)^{k-1}
\left(
1-\frac{\alpha \kav}{k^2_{\rm max} p(q)}\sum_{q''}q''W(q,q'')
\right)^{q-k}
\nonumber\\
&&\hspace*{-1cm}\times
\left(
\begin{array}{l}
q'-1\\
k'-1
\end{array}
\right)
\left(
\frac{\alpha \kav}{k^2_{\rm max}p(q)}\sum_{q''} q''W(q',q'')
\right)^{k'-1}
\left(
1-\frac{\alpha \kav}{k^2_{\rm max} p(q')}\sum_{q''} q''W(q',q'')
\right)^{q'-k'}
\eea

\item {\it Biased bond oversampling}, i.e. $x(k)=1$, $y(k,k^\prime)=1$, $z(k,k^\prime)=\alpha k k^\prime/k^2_{\rm max}$
\\[2mm]
Here we get
\begin{eqnarray}
&&
a(q)= \frac{\alpha}{k^2_{\rm max}}q \kav,~~~~~~~~~ b(q)=1
\\
&&
{\cal J}(k|q)=\frac{(\alpha \kav q/k^2_{\rm max})^{k-q}}{(k-q)!}e^{-\alpha \kav q/k^2_{\rm max}} I(k \geq q)
\\
&&
{\cal L}(k|q)=\frac{(\alpha \kav q/k^2_{\rm max})^{k-1-q}}{(k-1-q)!}e^{-\alpha \kav q/k^2_{\rm max}} I(k \geq q+1)
\end{eqnarray}
Hence we obtain
\bea
&&\hspace*{-2cm}
W(k,k'|x)=\frac{1}{\kav+\alpha \kav^2/k^2_{\rm max}}\sum_{qq'}e^{-\frac{\alpha \kav q}{k^2_{\rm max}}-\frac{\alpha \kav q'}{k^2_{\rm max}}}
\left[
\frac{\kav W(qq')}{(k-q)!(k'-q')!}\left(
\frac{\alpha \kav q}{k^2_{\rm max}}
\right)^{k-q}
\left(
\frac{\alpha \kav q'}{k^2_{\rm max}}
\right)^{k'-q'}
\right.\nonumber\\
&&\hspace*{-1cm}+\left.\frac{\alpha}{k^2_{\rm max}}\frac{p(q)p(q') q q'}{(k-q-1)!(k'-q'-1)!}
\left(
\frac{\alpha \kav q}{k^2_{\rm max}}
\right)^{k-q}
\left(
\frac{\alpha \kav q'}{k^2_{\rm max}}
\right)^{k'-q'}
\right]
\eea
\end{itemize}

\begin{figure}[t]
 \unitlength=0.31mm
\hspace*{-13mm}
\begin{picture}(200,590)
\put(50,400){\includegraphics[width=200\unitlength,angle=0]{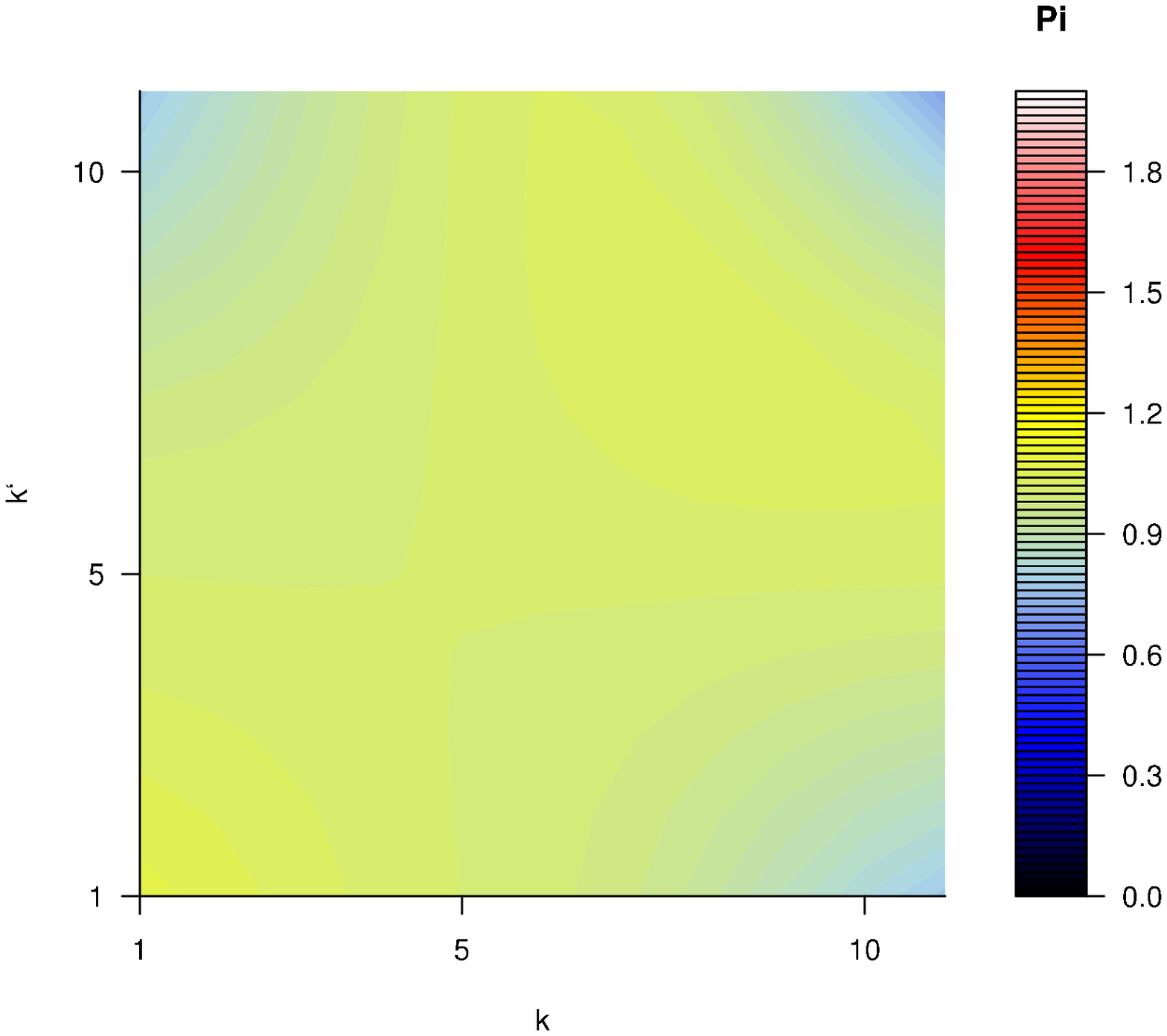}}
\put(270,400){\includegraphics[width=200\unitlength,angle=0]{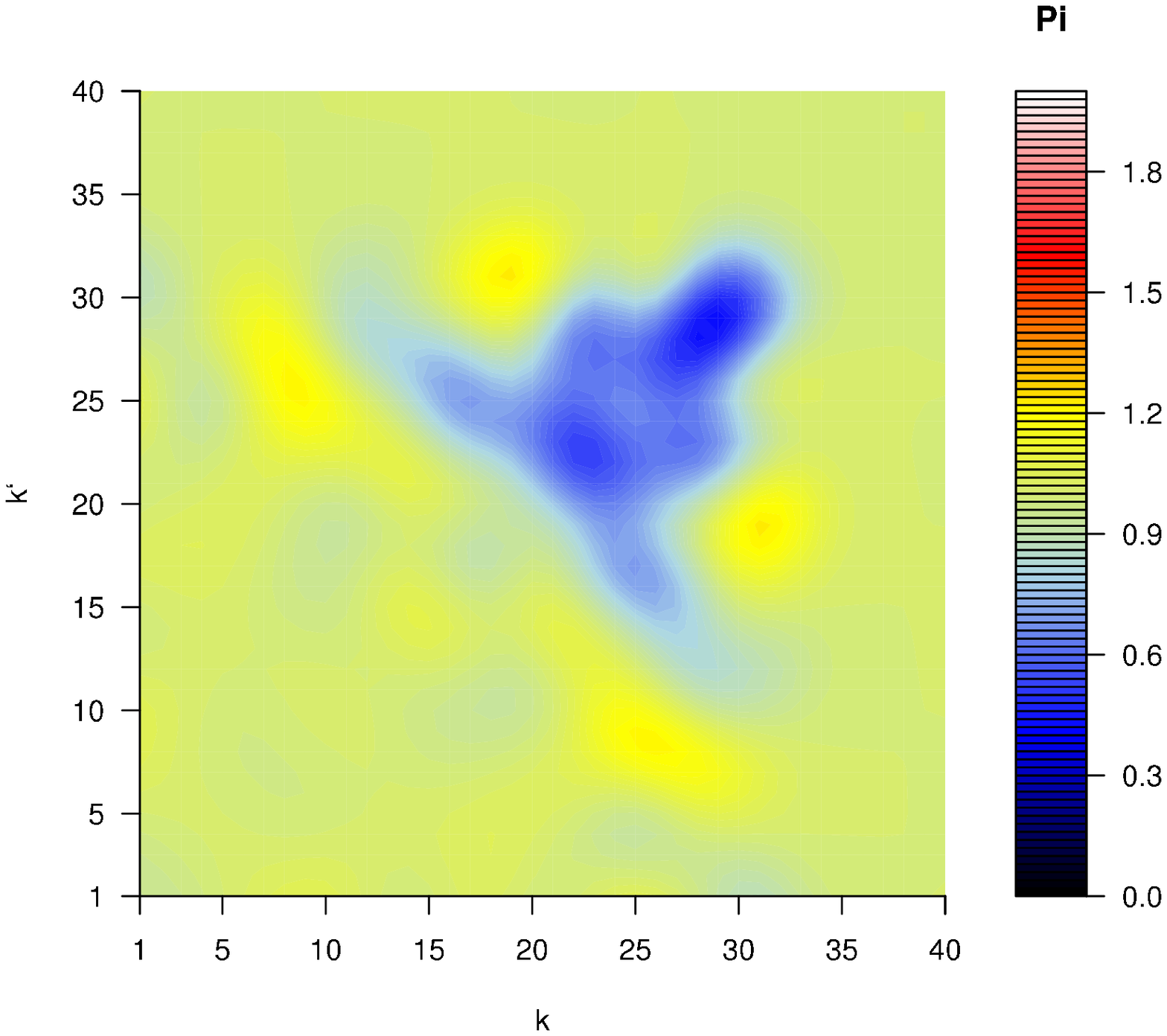}}
\put(50,200){\includegraphics[width=200\unitlength,angle=0]{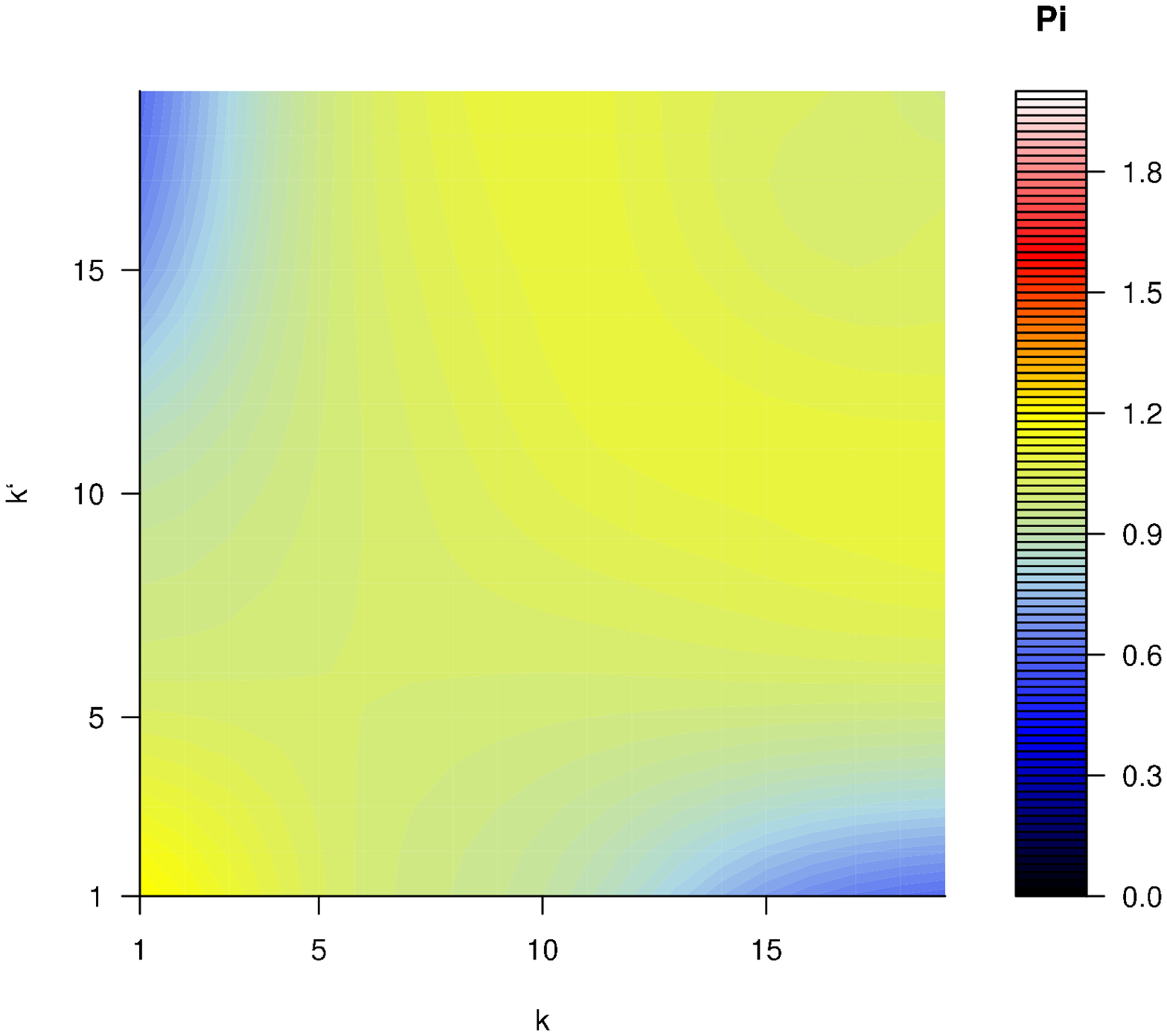}}
\put(270,200){\includegraphics[width=200\unitlength,angle=0]{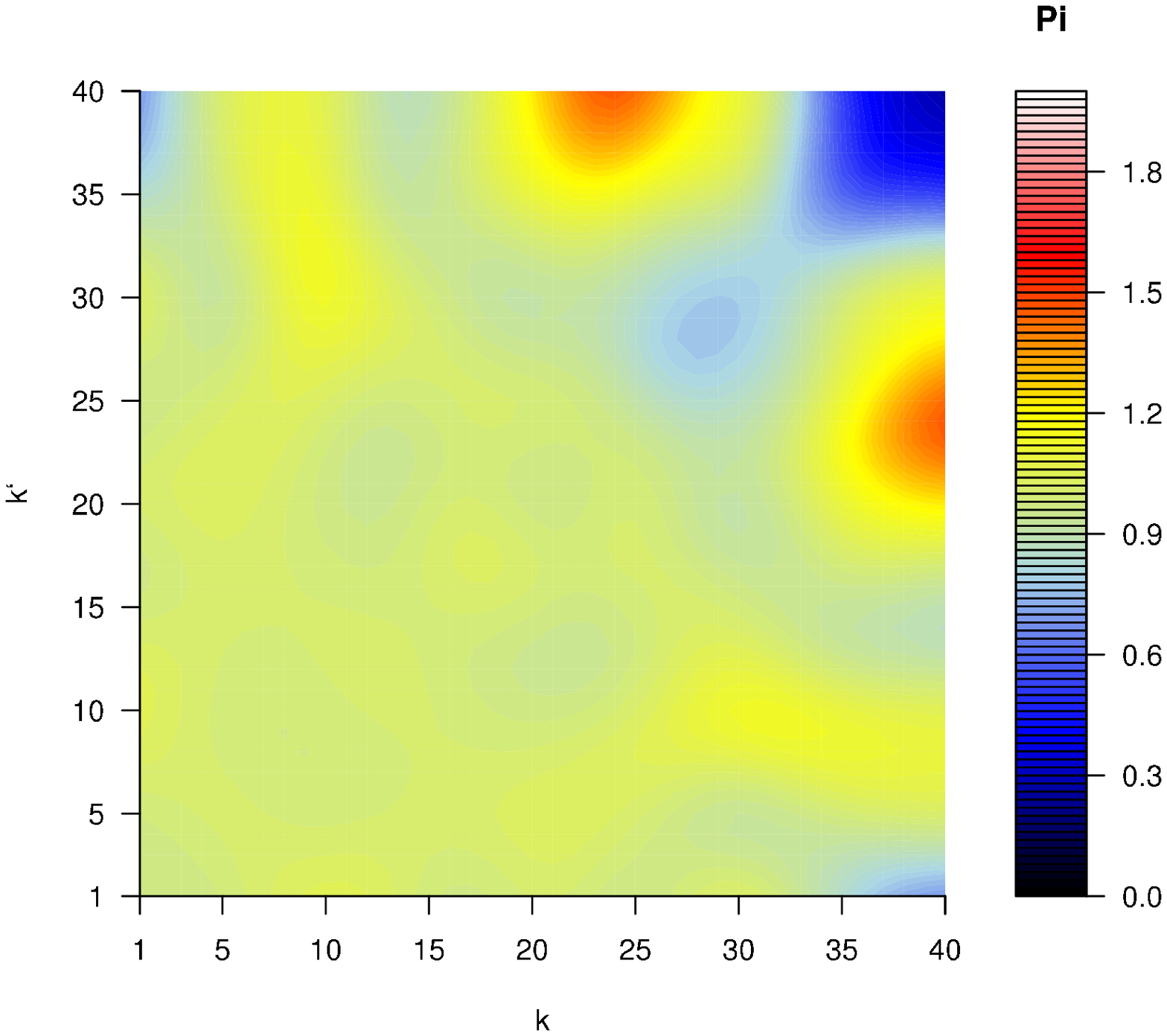}}
\put(50,10){\includegraphics[width=200\unitlength,angle=0]{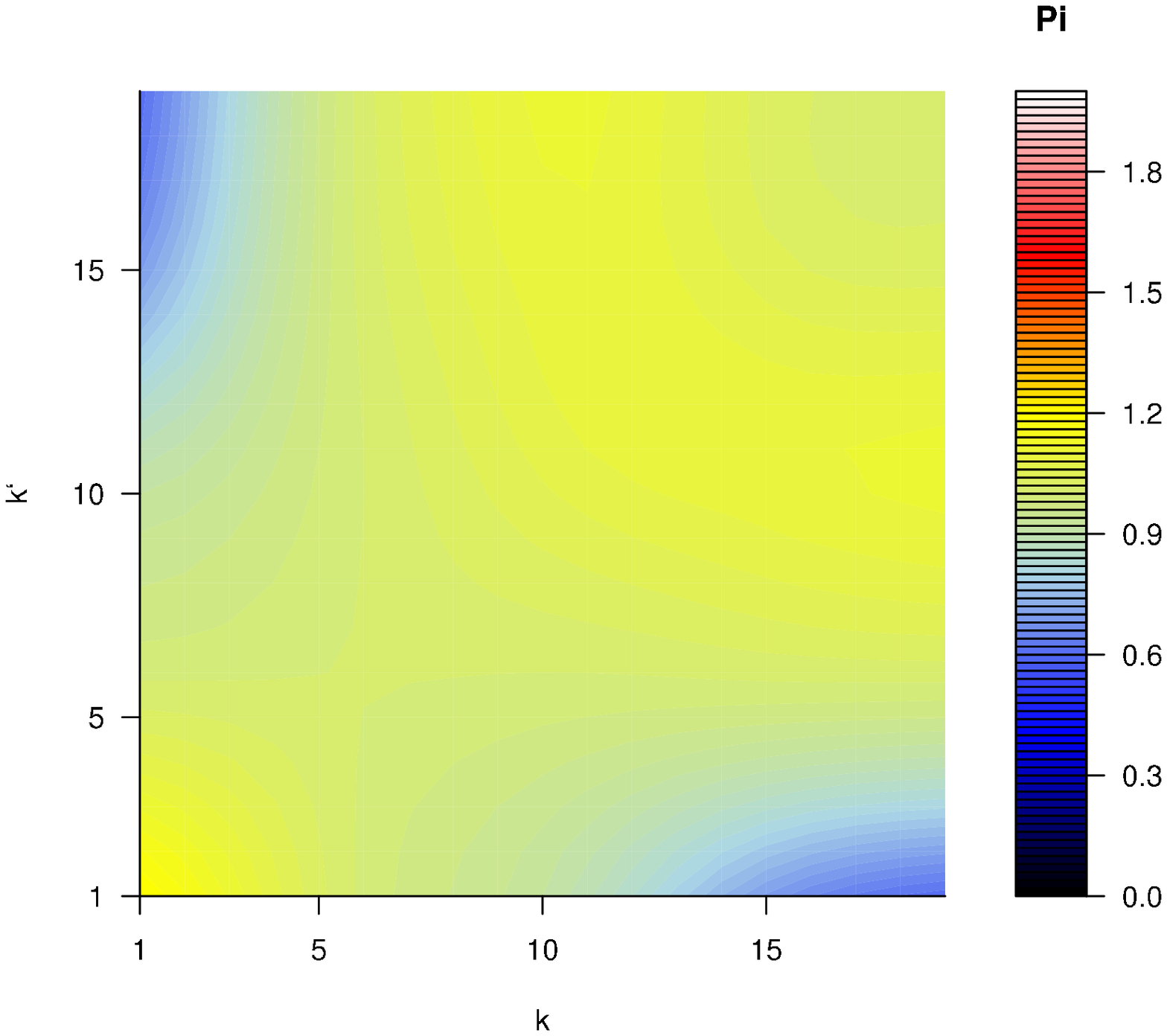}}
\put(270,10){\includegraphics[width=200\unitlength,angle=0]{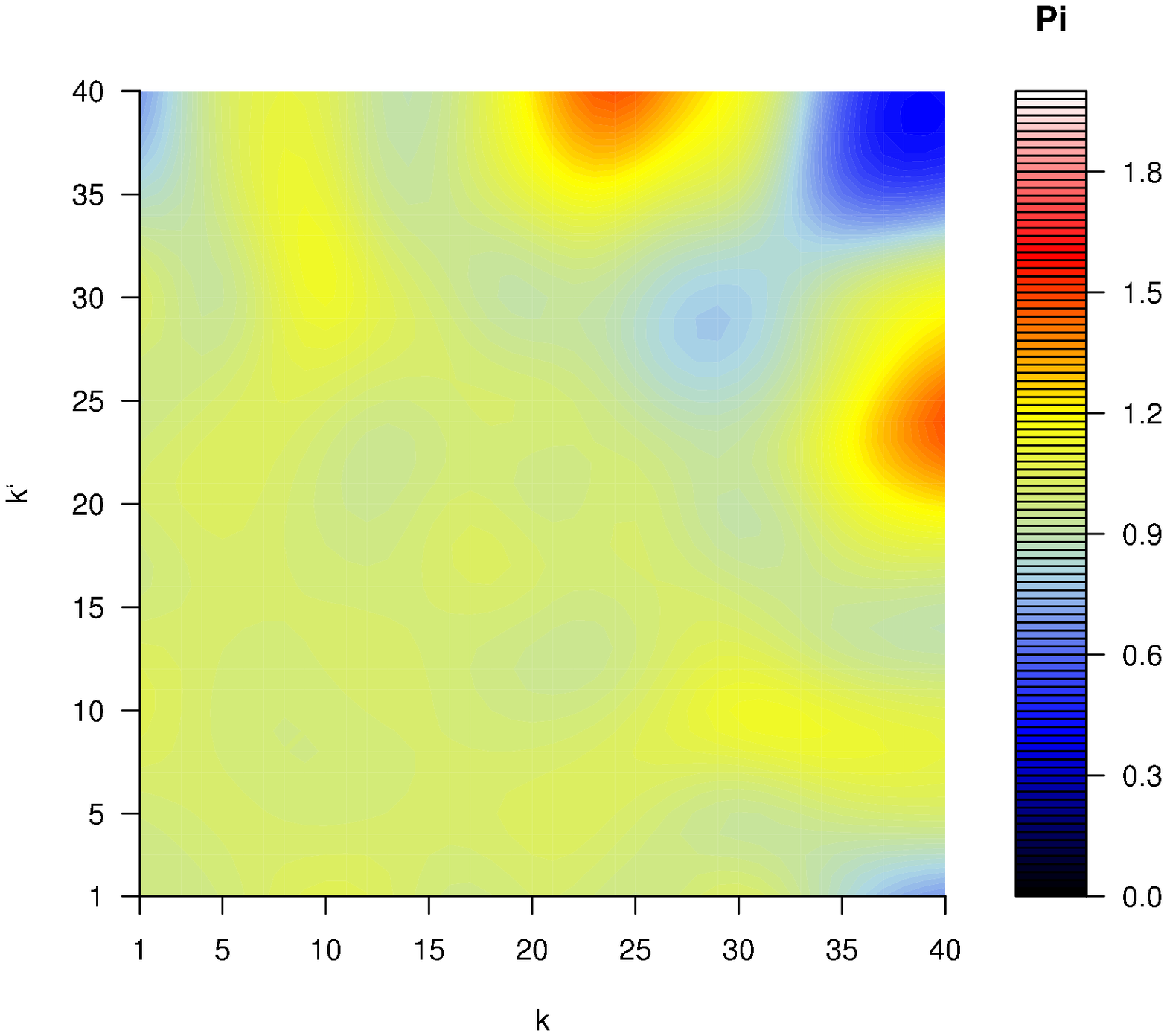}}
\end{picture}
\vspace*{-5mm}
\caption{
Effects of biased bond oversampling on the normalised degree correlation $\Pi(k,k^\prime)$
of the synthetically generated Poissonian
and Power law graphs ($N=3512$, $\kav=3.72$) shown in the top left and right panels, respectively. 
Middle panels show the result of simulations for $\alpha \kav^2/k^2_{\rm max}=1$ (left) and $\alpha \kav^2/k^2_{\rm max}=0.7$ (right)
and bottom panels show the corresponding theoretical predictions. Numerical data result from averaging over 
$10^4$ samples.
}
\end{figure}

\subsection{Summary}

As was the case for the degree distribution, also the degree correlations can for a broad and class of sampling protocols 
be calculated exactly and in terms of fully explicit relations. In contrast to the degree distribution, for which the sampling problem 
had already been studied partly by other authors,  
we are not aware of any analytical results for degree correlations. Our equations revealed that: 
\begin{itemize}
\item Sampling will always affect the degree correlations of networks, even in the unbiased case, 
if the original networks had such degree correlations. 
\item Uncorrelated networks will remain uncorrelated after sampling only for unbiased node and/or 
bond undersampling. Bond oversampling will in general introduce degree correlations, even in the 
unbiased case.
\item Unbiased node and bond undersampling both modify the degree correlations (and the degree distribution) 
in the same way, so they are equivalent for any graph with prescribed topological features $p(k)$ and 
$W(k,k')$, as generated from (\ref{eq:ensemble}).
\item Node and bond undersampling cannot be mapped onto each other in the case of biased sampling; their effects are qualitatively different. 
\end{itemize}

\section{Discussion}

It is well known that the presently available data on cellular signalling networks are incomplete, and often suffer from serious experimental bias, reflecting the highly nontrivial nature of the experimental methods available for their collection.
Yet a significant number of research papers continue to be written in which such data are used to infer
statements on the possible biological relevance of local network modules or motifs. In addition the signalling network data  are increasingly used for preprocessing gene expression data in order to derive more robust disease specific prognostic signatures \cite{Chuang,Rapaport,Taylor}, and will very soon impact on actual treatment decisions in medicine (e.g. will be used to suggest which cancer patients are likely to benefit from which chemotherapy).  
 Given this situation, it is vital that we understand quantitatively the data imperfections, i.e. 
the relation between the true biological signalling networks probed and the imperfect network samples of these networks 
that are reported in public data repositories and presently used by biomedical scientists. To do this we need mathematical tools; the relevant networks are too large to rely on numerical simulation alone. Moreover, unlike simulations, 
analytical results can be used in reverse to infer the most probable true networks from the imperfect observed samples. 

Ensembles of tailored random graphs with controlled topological properties 
are a natural and rigorous language for describing biological networks. 
They suggest precise definitions of structural features, they allow us to classify networks and obtain precise (dis)similarity measures, they provide precise `null models' for hypothesis testing, and they can serve as efficient proxies for real networks in process modelling.
In this paper we have shown how they can also be used to study analytically the effects of sampling on macroscopic topological properties of large biological networks, under a much wider range of conditions than those considered in previous analytical studies (the latter are recovered as special simple cases). 
We have obtained explicit expressions for both degree distributions and degree correlation kernels of sampled networks, and have been able to do this for  sampling protocols that involve node and/or link  undersampling as well as for link oversampling. 
Our predictions are in excellent agreement with numerical simulations. 

As could have been expected, the most dangerous types of sampling are the biased ones, where the probability to observe bonds or links depend on the degrees of the nodes concerned. 
Unfortunately, present experimental protocols are quite likely to involve precisely such sampling. We therefore 
hope that our new analytical tools, which take the form of explicit and transparent equations that  connect the topological structure functions $p(k)$ and $W(k,k^\prime)$ of the sampled and the true networks,  
can prove useful in explaining and decontaminating signalling network data.

\section*{Acknowledgements}

The authors are grateful to L Fernandes, F Fraternali, J Kleinjung and ES Roberts for stimulating discussions.   
ACCC would like to thank the Engineering and Physical Sciences Research Council (UK) and the Biotechnology and Biological Sciences Research Council (UK) for support.

\clearpage
\appendix

\section{Joint degree distribution of connected nodes}
\label{app:generalW}

\subsection{Path integral representation of $W(k,k^\prime|x,y,z)$}

Here we calculate the joint degree distribution of connected nodes (\ref{eq:W_tofind}) that will be observed in large networks that are sampled, according to protocol  (\ref{eq:cprime}), from 
typical graphs with prescribed macroscopic topological features  $p(k)$ and $W(k,k^\prime)$, as generated from (\ref{eq:ensemble}). With the short-hands
$\tilde{W}(k,k^\prime|x,y,z)=W(k,k^\prime|x,y,z)\bar{k}(x,y,z)\sum_{q} p(q)x(q)$, $\bk=(k_1,\ldots,k_N)$, and $\bOmega=(\Omega_1,\ldots,\Omega_N)\in[-\pi,\pi]^N$ we may write
\begin{eqnarray}
\hspace*{-20mm}
\tilde{W}(k,k^\prime|x,y,z)&=& \lim_{N\to\infty}\sum_{\bc}p(\bc)\Big\bra
\frac{1}{N}\sum_{ij} c^\prime_{ij}\delta_{k,\sum_\ell c^\prime_{i\ell}}
\delta_{k^\prime,\sum_\ell c^\prime_{j\ell}}\Big\ket_{\!\bsigma,\btau,\blambda}
\nonumber
\\
\hspace*{-20mm}
&=& 
\int_{\pi}^\pi\!\frac{\rmd\omega\rmd\omega^\prime}{4\pi^2}\rme^{\rmi(\omega k+\omega^\prime k^\prime)}
\lim_{N\to\infty}\frac{1}{Z_N}\int\!\frac{\rmd\bOmega}{(2\pi)^N}\rme^{\rmi\bOmega\cdot\bk}
\nonumber
\\
&&\times \sum_{\bc}
\prod_{r<s}
\Big[\delta_{c_{rs},1}\frac{\overline{k}}{N}\frac{W(k_r,k_s)}{p(k_r)p(k_s)}\rme^{-\rmi(\Omega_r+\Omega_s)}
\!+\!\delta_{c_{rs},0}(1\!-\!\frac{\overline{k}}{N}\frac{W(k_r,k_s)}{p(k_r)p(k_s)})
\Big]
\nonumber
\\
&&\times\frac{1}{N}\sum_{ij}
\Big\bra
 c^\prime_{ij}\rme^{-\rmi(\omega+\omega^\prime)-\rmi\omega\sum_{\ell\neq i,j} c^\prime_{i\ell}
-\rmi\omega^\prime \sum_{\ell\neq i,j} c^\prime_{j\ell}}\Big\ket_{\!\bsigma,\btau,\blambda}
\nonumber
\\
\hspace*{-20mm}
&=& 
\int_{\pi}^\pi\!\frac{\rmd\omega\rmd\omega^\prime}{4\pi^2}\rme^{\rmi(\omega (k-1)+\omega^\prime (k^\prime-1))}
\lim_{N\to\infty}\frac{1}{Z_N}\int\!\frac{\rmd\bOmega}{(2\pi)^N}\rme^{\rmi\bOmega\cdot\bk}\frac{1}{N}\sum_{ij}x(k_i)x(k_j)
\nonumber
\\
&&\times \sum_{\bc}
\prod_{r<s}
\Big[\delta_{c_{rs},1}\frac{\overline{k}}{N}\frac{W(k_r,k_s)}{p(k_r)p(k_s)}\rme^{-\rmi(\Omega_r+\Omega_s)}
\!+\!\delta_{c_{rs},0}(1\!-\!\frac{\overline{k}}{N}\frac{W(k_r,k_s)}{p(k_r)p(k_s)})
\Big]
\nonumber
\\
&&\times
\Big\bra
[\tau_{ij}c_{ij}\!+\!(1\!-\!c_{ij})\lambda_{ij}]\rme^{-\rmi\omega\sum_{\ell\neq i,j} \sigma_\ell (\tau_{i\ell}c_{i\ell}+\lambda_{i\ell}(1-c_{i\ell}))
-\rmi\omega^\prime \sum_{\ell\neq i,j} \sigma_\ell(\tau_{j\ell}c_{j\ell}+\lambda_{j\ell}(1-c_{j\ell}))}\Big\ket_{\!\bsigma,\btau,\blambda}
\nonumber
\\
\hspace*{-20mm}
&=& 
\int_{\pi}^\pi\!\frac{\rmd\omega\rmd\omega^\prime}{4\pi^2}\rme^{\rmi(\omega (k-1)+\omega^\prime (k^\prime-1))}
\lim_{N\to\infty}\frac{1}{Z_N}\rme^{\sum_{r<s}\log\Big[1+\frac{\overline{k}}{N}\frac{W(k_r,k_s)}{p(k_r)p(k_s)}\Big(\rme^{-\rmi(\Omega_r+\Omega_s)}\!-\!1\Big)
\Big]}
\nonumber
\\
&&\times\int\!\frac{\rmd\bOmega}{(2\pi)^N}\rme^{\rmi\bOmega\cdot\bk}\frac{1}{N^2}\sum_{ij}x(k_i)x(k_j)
\Big[z(k_i,k_j)\!+\!\overline{k}y(k_i,k_j)\frac{W(k_i,k_j)}{p(k_i)p(k_j)}\rme^{-\rmi(\Omega_i+\Omega_j)}
\Big]
\nonumber
\\
&&
\times
\rme^{(\rme^{-\rmi\omega}-1)
\frac{1}{N}\sum_\ell x(k_\ell)\Big[z(k_i,k_\ell)+\overline{k}y(k_i,k_\ell)\frac{W(k_i,k_\ell)}{p(k_i)p(k_\ell)}\rme^{-\rmi(\Omega_i+\Omega_\ell)}\Big]}
\nonumber
\\
&&
\times
\rme^{(\rme^{-\rmi\omega^\prime}-1)
\frac{1}{N}\sum_\ell x(k_\ell)\Big[z(k_j,k_\ell)+\overline{k}y(k_j,k_\ell)\frac{W(k_j,k_\ell)}{p(k_j)p(k_\ell)}\rme^{-\rmi(\Omega_j+\Omega_\ell)}\Big]}
\label{eq:Wintermediate}
\end{eqnarray}
We next introduce the following order parameters 
\begin{eqnarray}
P(q,\Omega|\bk,\bOmega)&=& \frac{1}{N}\sum_r \delta_{q,k_r}\delta(\Omega-\Omega_r)
\end{eqnarray}
and insert into (\ref{eq:Wintermediate}) for each $(q,\Omega)$ the following integral:
\begin{eqnarray}
1&=& \int\!\rmd P(q,\Omega)~\delta\Big[P(q,\Omega)\!-\!P(q,\Omega|\bk,\bOmega)\Big]
\nonumber
\\
&=&(N/2\pi)\int\!\rmd P(q,\Omega)\rmd\hat{P}(q,\Omega)\rme^{\rmi N\hat{P}(q,\Omega)P(q,\Omega)-
\rmi\sum_r \delta_{q,k_r}\delta(\Omega-\Omega_r)}
\end{eqnarray}
This converts (\ref{eq:Wintermediate}) into the following path integral, with the short-hand $\{\rmd P\rmd\hat{P}\}=\prod_{q,\Omega}[\rmd P(q,\Omega)\rmd\hat{P}(q,\Omega)/2\pi]$ and with $Z^\prime_N$ a new constant that apart from containing $Z_N$ absorbs various factors $N$ and constants that are generated when transforming sums over $\Omega$ into integrals:
\begin{eqnarray}
\hspace*{-20mm}
\tilde{W}(k,k^\prime|x,y,z)&=&  \lim_{N\to\infty}\int\!\frac{\{\rmd P\rmd \hat{P}\}}
{Z_N^\prime}\rme^{N\Psi[P,\hat{P}]+\Phi[P,\hat{P}]+\order(N^{-1})}
\sum_{q,q^\prime}\int\!\rmd\Omega\rmd\Omega^\prime P(q,\Omega)P(q^\prime\!,\Omega^\prime)
x(q)x(q^\prime)
\nonumber
\\
\hspace*{-20mm}
&&\hspace*{-10mm}\times \Xi(q,\Omega;q^\prime\!,\Omega^\prime)
\Big(
\int_{\pi}^\pi\!\frac{\rmd\omega}{2\pi}\rme^{\rmi\omega (k-1)
+(\rme^{-\rmi\omega}-1)Q(q,\Omega)}
\Big)
\Big(\int_{\pi}^\pi\!\frac{\rmd\omega}{2\pi}
\rme^{\rmi\omega(k^\prime-1)
+(\rme^{-\rmi\omega}\!-1)Q(q^\prime,\Omega^\prime)
}
\Big)
\label{eq:Wintermediate2}
\end{eqnarray}
in which $\Phi[P,\hat{P}]$ will eventually drop out of our formulae (via normalization) and 
\begin{eqnarray}
\Psi[P,\hat{P}]&=& \rmi\sum_q\int\!\rmd\Omega~\hat{P}(q,\Omega)P(q,\Omega)+\sum_k p(k) \log \int\!\frac{\rmd\Omega}{2\pi}\rme^{\rmi\Omega k-\rmi\hat{P}(k,\Omega)}
\nonumber
\\
&&+
\frac{1}{2}\overline{k}\sum_{qq^\prime}\int\!\rmd\Omega\rmd\Omega^\prime~P(q,\Omega)P(q^\prime\!,\Omega^\prime)
\frac{W(q,q^\prime)}{p(q)p(q^\prime)}(\rme^{-\rmi(\Omega+\Omega^\prime)}-1)
\label{eq:Psi}
\\
\Xi(q,\omega;q^\prime,\omega^\prime)&=&
z(q,q^\prime)+y(q,q^\prime)
\overline{k}\frac{W(q,q^\prime)}{p(q)p(q^\prime)}\rme^{-\rmi(\omega+\omega^\prime)}
\\
Q(q,\Omega)&=& 
\sum_{q^\pprime}\!\int\!\rmd\Omega^\pprime P(q^\pprime,\Omega^\pprime)x(q^\pprime)~
\Xi(q,\Omega;q^\pprime\!,\Omega^\pprime)
\label{eq:Q}
\end{eqnarray}
The relevant $\omega$-integrals are of the familiar form
\begin{eqnarray}
\int_{\pi}^\pi\!\frac{\rmd\omega}{2\pi}\rme^{\rmi\omega \ell
+(\rme^{-\rmi\omega}-1)Q}=\rme^{-Q}\sum_{n\geq 0}\frac{Q^n}{n!}
 \int_{\pi}^\pi\!\frac{\rmd\omega}{2\pi}\rme^{\rmi\omega (\ell-n)}=\rme^{-Q}Q^\ell/\ell!
 \label{eq:Iintegral}
\end{eqnarray}
(unless $\ell<0$, in which case the integral is zero). 
We also note that by definition we always have the normalisation identity $\sum_{k,k^\prime\geq 0}W(k,k^\prime|x,y,z)=1$.  
So we arrive at:
\begin{eqnarray}
&&
\hspace*{-23mm}
W(k,k^\prime|x,y,z)= \notdelta_{k,0}\notdelta_{k^\prime,0}\frac{
\sum_{q,q^\prime}x(q)x(q^\prime)\int\!\rmd\Omega\rmd\Omega^\prime P(q,\Omega)P(q^\prime\!,\Omega^\prime)
 \Xi(q,\Omega;q^\prime\!,\Omega^\prime)I(k|q,\Omega)I(k^\prime|q^\prime,\Omega^\prime)
 }
  {
\sum_{q,q^\prime}x(q)x(q^\prime)\int\!\rmd\Omega\rmd\Omega^\prime P(q,\Omega)P(q^\prime\!,\Omega^\prime)
 \Xi(q,\Omega;q^\prime\!,\Omega^\prime)
}\nonumber
\\
&&\label{eq:Wintermediate3}
\end{eqnarray}
\vspace*{-9mm}

\noindent
with 
\begin{eqnarray}
I(k|q,\Omega)=\rme^{-Q(q,\Omega)}Q^{k-1}(q,\Omega)/(k\!-\!1)!
\label{eq:I}
\end{eqnarray}
 and 
in which, via the steepest descent argument, the order parameters $\{P,\hat{P}\}$ are the functions that extremise the kernel (\ref{eq:Psi}). 

\subsection{Functional saddle-point equations}

Functional variation of (\ref{eq:Psi}) gives the following saddle-point equations for $\{P,\hat{P}\}$:
\begin{eqnarray}
&& \rmi\hat{P}(q,\Omega)=-\overline{k}\sum_{q^\prime}\int\!\rmd\Omega^\prime~P(q^\prime\!,\Omega^\prime)
\frac{W(q,q^\prime)}{p(q)p(q^\prime)}(\rme^{-\rmi(\Omega+\Omega^\prime)}-1)
\label{eq:spe1}
\\
&&  P(q,\Omega)= p(q) \frac{\rme^{\rmi\Omega q-\rmi\hat{P}(q,\Omega)}}{ \int\!\rmd\Omega^\prime ~
\rme^{\rmi\Omega^\prime q-\rmi\hat{P}(q,\Omega^\prime)}}
\label{eq:spe2}
\end{eqnarray}
Equivalently:
\begin{eqnarray}
&&
\rmi\hat{P}(q,\Omega)=\overline{k}\lambda(q)-\overline{k}\rme^{-\rmi\Omega}\phi(q),~~~~~~~~
P(q,\Omega)= p(q) \frac{\rme^{\rmi\Omega q+\overline{k}\rme^{-\rmi\Omega}\phi(q)}}{ \int\!\rmd\Omega^\prime ~
\rme^{\rmi\Omega^\prime q+\overline{k}\rme^{-\rmi\Omega^\prime}\phi(q)}}
\label{eq:hatPP}
\end{eqnarray}
with
\begin{eqnarray}
\phi(q)&=&\sum_{q^\prime}
\frac{W(q,q^\prime)}{p(q)p(q^\prime)}\int\!\rmd\Omega~P(q^\prime\!,\Omega)\rme^{-\rmi \Omega}
\label{eq:phi}
\\
\lambda(q)&=& \sum_{q^\prime}
\frac{W(q,q^\prime)}{p(q)p(q^\prime)}\int\!\rmd\Omega~P(q^\prime\!,\Omega)
\label{eq:lambda}
 \end{eqnarray}
The integrals over $\Omega$ in (\ref{eq:phi}) and (\ref{eq:lambda}) are again of the type (\ref{eq:Iintegral}), 
from which we derive $\int\!\rmd\Omega~P(q,\Omega)=p(q)$ and $\int\!\rmd\Omega~P(q,\Omega)\rme^{-\rmi\Omega}=p(q)q/\overline{k}\phi(q)$. 
This then converts (\ref{eq:phi},\ref{eq:lambda}) into 
\begin{eqnarray}
\phi(q)p(q)&=&\sum_{q^\prime}
W(q,q^\prime)q^\prime/\overline{k}\phi(q^\prime),~~~~~~~~
\lambda(q)= \frac{1}{p(q)}\sum_{q^\prime}W(q,q^\prime)
\label{eq:philambda}
 \end{eqnarray}
 Since we know the marginal of the distribution $W(k,k^\prime)$ to be $\sum_{k^\prime}W(k,k^\prime)=kp(k)/\overline{k}$ (which follows directly from its definition), we can immediately read off the solution of (\ref{eq:philambda}):
 \begin{eqnarray}
 \phi(q)=\lambda(q)=q/\overline{k}
 \end{eqnarray}
 Insertion into  (\ref{eq:hatPP}) and using (\ref{eq:Iintegral}) gives the solution of (\ref{eq:spe1},\ref{eq:spe2}) in explicit form:
 \begin{eqnarray}
&&
\rmi\hat{P}(q,\Omega)=q-q\rme^{-\rmi\Omega},~~~~~~~~
P(q,\Omega)= p(q) \frac{\rme^{\rmi\Omega q+q\rme^{-\rmi\Omega}}}{
2\pi q^q/q!}
\end{eqnarray}

\subsection{Final result for the distribution $W(k,k^\prime|x,y,z)$}

We can now evaluate the various ingredients of (\ref{eq:Wintermediate3}). The function $Q(q,\Omega)$ becomes
\begin{eqnarray}
Q(q,\Omega)&=& 
\sum_{q^\prime\geq 0}
p(q^\prime)x(q^\prime)z(q,q^\prime)
+
\overline{k}\rme^{-\rmi\Omega}\frac{1}{p(q)}\sum_{q^\prime\geq 0}
x(q^\prime)y(q,q^\prime)W(q,q^\prime)
\label{eq:Qfinal}
\end{eqnarray}
Hence
\begin{eqnarray}
&&\hspace*{-15mm}
\int\!\rmd\Omega\rmd\Omega^\prime P(q,\Omega)P(q^\prime\!,\Omega^\prime)
 \Xi(q,\Omega;q^\prime\!,\Omega^\prime)I(k|q,\Omega)I(k^\prime|q^\prime,\Omega^\prime)
\nonumber
\\
&=&
\frac{p(q)q!}{q^q(k\!-\!1)!}\int\!\frac{\rmd\Omega}{2\pi}~\rme^{\rmi\Omega q+q\rme^{-\rmi\Omega}-Q(q,\Omega)}Q^{k-1}(q,\Omega)
\nonumber
\\
&&\times
\frac{p(q^\prime)(q^\prime)!}{(q^\prime)^{q^\prime}(k^\prime\!-\!1)!}
\int\!\frac{\rmd\Omega^\prime}{2\pi}~ 
\rme^{\rmi\Omega^\prime q^\prime+q^\prime\rme^{-\rmi\Omega^\prime}-Q(q^\prime,\Omega^\prime)}Q^{k^\prime-1}(q^\prime,\Omega^\prime)
\nonumber
\\
&&\times
\Big[ z(q,q^\prime)+y(q,q^\prime)
\overline{k}\frac{W(q,q^\prime)}{p(q)p(q^\prime)}\rme^{-\rmi(\Omega+\Omega^\prime)}\Big]
\nonumber
\\
&=&
 p(q)p(q^\prime)z(q,q^\prime){\cal J}(k|q){\cal J}(k^\prime|q^\prime)
+\overline{k}W(q,q^\prime)y(q,q^\prime)
{\cal L}(k|q){\cal L}(k^\prime|q^\prime)
\end{eqnarray}
in which 
\begin{eqnarray}
{\cal J}(k|q)&=& \notdelta_{k,0}\frac{q!}{q^q (k\!-\!1)!}\int_{-\pi}^\pi\!\frac{\rmd\Omega}{2\pi}~ \rme^{\rmi\Omega q+q\rme^{-\rmi\Omega}}
Q^{k-1}(q,\Omega)\rme^{-Q(q,\Omega)}
\label{eq:Js}
\end{eqnarray}
and
\begin{eqnarray}
{\cal L}(k|q)&=& \notdelta_{k,0}\frac{q!}{q^q (k\!-\!1)!}\int_{-\pi}^\pi\!\frac{\rmd\Omega}{2\pi}~ \rme^{\rmi\Omega (q-1)+q\rme^{-\rmi\Omega}}
Q^{k-1}(q,\Omega)\rme^{-Q(q,\Omega)}
\label{eq:Ls}
\end{eqnarray}
Summation over $k$ reveals that $\sum_{k\geq 0}{\cal J}(k|q)=\sum_{k\geq 0}{\cal L}(k|q)=1$ for all $q>0$, which 
leads to the final result:
\begin{eqnarray}
&&\hspace*{-23mm}
W(k,k^\prime|x,y,z)\nonumber
\\
\hspace*{-23mm}
&&\hspace*{-17mm} =
\frac{\sum_{q,q^\prime>0}x(q)x(q^\prime)\Big\{p(q)p(q^\prime)z(q,q^\prime){\cal J}(k|q){\cal J}(k^\prime|q^\prime)
+\overline{k}W(q,q^\prime)y(q,q^\prime)
{\cal L}(k|q){\cal L}(k^\prime|q^\prime)\Big\}}
{\sum_{q,q^\prime>0}x(q)x(q^\prime)\Big\{p(q)p(q^\prime)z(q,q^\prime)
+\overline{k}W(q,q^\prime)y(q,q^\prime)
\Big\}}
\nonumber
\\
\hspace*{-23mm}
&&\hspace*{-17mm} =
\frac{\sum_{q,q^\prime>0}x(q)x(q^\prime)\Big\{p(q)p(q^\prime)z(q,q^\prime){\cal J}(k|q){\cal J}(k^\prime|q^\prime)
+\overline{k}W(q,q^\prime)y(q,q^\prime)
{\cal L}(k|q){\cal L}(k^\prime|q^\prime)\Big\}}
{\bar{k}(x,y,z)\sum_q p(q)x(q)}
\nonumber
\\[-1mm]&&\label{eq:fullW}
\end{eqnarray}
with $\overline{k}(x,y,z)$ as given in (\ref{eq:average_degree}). The marginals of $W(k,k^\prime|x,y,z)$ are obtained trivially by summing (\ref{eq:fullW}) over $k^\prime$, giving 
\begin{eqnarray}
&&\hspace*{-23mm}
W(k|x,y,z)
= \frac{\sum_{q,q^\prime>0}x(q)x(q^\prime)\Big\{p(q)p(q^\prime)z(q,q^\prime){\cal J}(k|q)
+\overline{k}W(q,q^\prime)y(q,q^\prime)
{\cal L}(k|q)\Big\}}
{\bar{k}(x,y,z)\sum_q p(q)x(q)}
\label{eq:Wmarginals}
\end{eqnarray}

\subsection{Explicit expression for the factors ${\cal J}(k|q)$}

To carry out the integral in (\ref{eq:Js},\ref{eq:Ls}) we first write 
$Q(q,\Omega)$ as $Q(q,\Omega)=a(q)+b(q)q\rme^{-\rmi\Omega}$, with
\begin{eqnarray}
&& a(q)=
\sum_{q^\prime\geq 0}
p(q^\prime)x(q^\prime)z(q,q^\prime),~~~~~~b(q)=
\frac{\overline{k}}{qp(q)}\sum_{q^\prime\geq 0}
x(q^\prime)y(q,q^\prime)W(q,q^\prime)
\end{eqnarray}
We note that, due to $\sum_{q^\prime}W(q,q^\prime)=(q/\overline{k})p(q)$, we can be sure that $a(q)\in[0,1]$ and $b(q)\in[0,1]$.
Substitution into (\ref{eq:Js},\ref{eq:Ls}) and integration over $\Omega$, 
for $q>0$ and $k>0$, then leads to 
\begin{eqnarray}
{\cal J}(k|q)&=& \rme^{-a(q)}\frac{q!}{q^q (k\!-\!1)!}
\sum_{n=0}^{k-1}\Big(\!\begin{array}{c}k\!-\!1\\ n\end{array}\!\Big)a^{k-1-n}(q)b^n(q)q^n
\int_{-\pi}^\pi\!\frac{\rmd\Omega}{2\pi}~ \rme^{\rmi\Omega (q-n)+q(1-b(q))\rme^{-\rmi\Omega}}
\nonumber
\\
&=&
\rme^{-a(q)}
\sum_{n=0}^{{\rm min}\{k-1,q\}}
\Big(\!\begin{array}{c}q\\ n\end{array}\!\Big)
 \frac{a^{k-1-n}(q)}{(k\!-\!1\!-\!n)!}
b^n(q)(1-b(q))^{q-n}
\end{eqnarray}
and, similarly, 
\begin{eqnarray}
{\cal L}(k|q)
=
\rme^{-a(q)}
\sum_{n=0}^{{\rm min}\{k-1,q-1\}}
\Big(\!\begin{array}{c}q-1\\ n\end{array}\!\Big)
 \frac{a^{k-1-n}(q)}{(k\!-\!1\!-\!n)!}
b^n(q)(1-b(q))^{q-1-n}
\end{eqnarray}
Clearly ${\cal J}(k|q)\geq 0, {\cal L}(k|q)\geq 0$ for all $(k,q)$. Since 
the factors (\ref{eq:Js},\ref{eq:Ls}) also satisfy the normalization 
$\sum_{k\geq 0}{\cal J}_{kq}=1, \sum_{k\geq 0}{\cal L}_{kq}=1$ for all $q>0$, 
they can be interpreted as conditional probabilities, 
as suggested by our chosen notation. 

\subsection{Tests}

To test our expression (\ref{eq:fullW})  we set $x(k)=x$, $y(k)=y$ and $z(k,k^\prime)=z$, and try to recover from  (\ref{eq:Wmarginals}) via (\ref{eq:connection}) our earlier results on the degree distribution for unbiased sampling. We now find 
$ a(q)=xz$, $b(q)=xy$, and $\bar{k}(x,y,z)=x (z+\bar{k} y)$, which implies that
\begin{eqnarray}
{\cal J}(k|q)&=& \rme^{-xz}x^{k-1}\sum_{n=0}^{{\rm min}\{q,k-1\}}\!
\Big(\!\begin{array}{c}q\\ n\end{array}\!\Big)~
y^n(1\!-\!xy)^{q-n}\frac{z^{k-1-n}}{(k\!-\!1\!-\!n)!}
\end{eqnarray}
and
\begin{eqnarray}
{\cal L}(k|q)&=& {\cal J}(k|q\!-1)
\end{eqnarray}
Let us inspect the following cases:
\begin{itemize}
\item
{\em Perfect sampling,} i.e. $x=y=1$ and $z=0$.
\\[2mm]
 Now there should be no difference between the kernel $W(k,k^\prime)$ and the observed kernel $W(k,k^\prime|1,1,0)$ of the sample. 
 Here we see that (\ref{eq:Qfinal}) simplifies to $Q(q,\Omega)=q\rme^{-\rmi\Omega}$; hence $a(q)=0$ and $b(q)=1$, leading to ${\cal L}(k|q)=\delta_{q,k}$ and therefore  to 
the correct identity $W(k,k^\prime|x,y,z)= W(k,k^\prime)$. 
\item
{\em Unbiased node and/or link undersampling}, i.e. $xy<1$ and $z=0$.
\\[2mm]
Now we have $\bar{k}(x,y,0)=\bar{k}xy$ and
\begin{eqnarray}
{\cal J}(k|q)&=& x^{k-1}
\Big(\!\begin{array}{c}q\\ k-1\end{array}\!\Big)
y^{k-1}(1\!-\!xy)^{q-k+1}
I(q\geq k-1)
\end{eqnarray}
which gives
\begin{eqnarray}
W(k|x,y,0)
&=& \frac{1}{\bar{k}}\sum_{k^\prime\geq k}p(k^\prime) k^\prime
\Big(\!\begin{array}{c}k^\prime\!-\!1\\ k\!-\!1\end{array}\!\Big)
(xy)^{k-1}(1\!-\!xy)^{k^\prime\!-\!k}
\end{eqnarray}
and therefore we recover the correct expression
\begin{eqnarray}
p(k|x,y,0)&=& \frac{\bar{k}xy}{k}W(k|x,y,0)=
(xy)^{k}\sum_{k^\prime\geq k}p(k^\prime) 
\Big(\!\begin{array}{c}k^\prime\\ k^\prime\!-\!k\end{array}\!\Big)
(1\!-\!xy)^{k^\prime\!-\!k}
\end{eqnarray}
\item
{\em Unbiased bond oversampling,} i.e. $x=y=1$ and $z>0$.
\\[2mm]
Now $\bar{k}(1,1,z)=\bar{k}+z$ and 
${\cal J}(k|q)=\rme^{-z}\frac{z^{k-1-q}}{(k-1-q)!}I(k\geq q\!+\!1)$, which results in 
\begin{eqnarray}
p(k|1,1,z)&=& \frac{\bar{k}}{k}W(k|1,1,z)=
\frac{1}{k}\Big\{z\sum_q p(q){\cal J}(k|q)+\sum_q p(q)q{\cal J}(k|q\!-\!1)\Big\}
\nonumber
\\
&=& \frac{\rme^{-z}}{k}\Big\{\sum_{q=0}^{k-1}p(q)\frac{z^{k-q}}{(k\!-\!1\!-\!q)!}+\sum_{q=1}^{k}p(q)q\frac{z^{k-q}}{(k\!-\!q)!}\Big\}
\nonumber
\\
&=& \rme^{-z}\sum_{\ell=0}^k p(k\!-\!\ell)z^\ell/\ell!
\end{eqnarray}
which is indeed the correct result identified earlier. 

\end{itemize}

\end{document}